\documentclass{aa}  

\usepackage{graphicx}
\usepackage{txfonts}
\usepackage{makecell}

\usepackage{multirow}

\usepackage[colorlinks=true,allcolors=blue]{hyperref}

\newcommand{\nii}{[N{\sc{ii}}]}

\newcommand{\oiii}{[O{\sc{iii}}]}
\newcommand{\ha}{H$\alpha$\xspace}

\newcommand{\hb}{H$\beta$\xspace}

\newcommand{\mgii}{Mg{\sc{ii}}\xspace}
\newcommand{\civ}{C{\sc{iv}}\xspace}

\newcommand{\Mbh}{M$_{\rm BH}$\xspace}

\newcommand{\kms}{${\rm km~s^{-1}}$\xspace}

\newcommand{\Feii}{Fe{\sc{ii}}\xspace}

\defcitealias{Vestergaard:2006}{VP06}
\defcitealias{Reines:2013}{R13}
\defcitealias{Dallabonta:2020}{DB20}
\defcitealias{Dallabonta:2025}{DB25}
\defcitealias{Greene:2005}{GH05}

\begin{document}

   \title{Doubling NIRSpec/IFS capability to calibrate the single-epoch black hole mass relation at high redshifts}

   \subtitle{}

   \author{ Eleonora Parlanti\inst{\ref{iNorm}} \fnmsep\thanks{eleonora.parlanti@sns.it}
\and
Bartolomeo Trefoloni \inst{\ref{iNorm},\ref{iarcetri}}
\and
Stefano Carniani \inst{\ref{iNorm}} 
\and
Francesco D'Eugenio  \inst{\ref{inst:kavli}, \ref{inst:cavendish}}
\and
Michele Perna \inst{\ref{inst:centroastrob}}
\and
Giulia Tozzi \inst{\ref{MPE}}
\and
Hannah \"Ubler \inst{\ref{MPE}}
\and
Giacomo Venturi \inst{\ref{iNorm}}
\and 
Sandra Zamora \inst{\ref{iNorm}}
          }

   \institute{         Scuola Normale Superiore, Piazza dei Cavalieri 7, I-56126 Pisa, Italy\label{iNorm}
   \and
   INAF -- Osservatorio Astrofisico di Arcetri, Largo Enrico Fermi 5, I-50125 Firenze, Italy\label{iarcetri}
         \and
    Centro de Astrobiolog\'{\i}a (CAB), CSIC-INTA, Ctra. de Ajalvir km 4, Torrej\'on de Ardoz, E-28850, Madrid, Spain \label{inst:centroastrob} 
    \and
    Kavli Institute for Cosmology, University of Cambridge, Madingley Road, Cambridge, CB3 0HA, UK \label{inst:kavli} 
    \and
    Cavendish Laboratory - Astrophysics Group, University of Cambridge, 19 JJ Thomson Avenue, Cambridge, CB3 0HE, UK \label{inst:cavendish} 
    \and
             Max-Planck-Institut f\"ur extraterrestrische Physik (MPE), Gie{\ss}enbachstra{\ss}e 1, 85748 Garching, Germany\label{MPE}
             }

   \date{}

   \titlerunning{Doubling NIRSpec/IFS capability}
\authorrunning{E. Parlanti et al.}

% \abstract{}{}{}{}{} 
% 5 {} token are mandatory

\abstract 
{
The recent discovery of a large population of overmassive black holes (BHs) in the early Universe, enabled by JWST, challenges the validity of the BH–host galaxy coevolution framework. However, the reliability of the estimated BH masses remains uncertain, as they are typically derived using single-epoch (SE) mass relations calibrated locally.
Calibrating SE relations directly in the high-redshift Universe would allow us to test their validity beyond the local Universe and assess any potential bias affecting BH mass estimates in the early Universe.

In this work, we present a data-reduction technique for JWST/NIRSpec Integral Field Unit (IFU) observations that doubles the effective wavelength coverage of medium-resolution gratings (G140M, G235M). This enables the detection of emission that would be inaccessible using a single grating and therefore requires observations with two adjacent gratings. This new pipeline offers observers the possibility of integrating longer with a single ``bluer'' configuration rather than splitting exposure time across adjacent gratings.
We applied this pipeline to a sample of five $z\sim2$ quasars with BH masses independently measured through reverberation mapping (RM) campaigns, and NIRSpec IFU observations of the H$\beta$ emission line with G140M/F100LP. Our pipeline allows us to recover the H$\alpha$ emission located well beyond the NIRSpec nominal wavelength range.
Using this sample, we assessed the reliability of the most widely adopted SE calibrations, finding that H$\beta$ yields the closest agreement with the BH masses estimated from RM, whereas H$\alpha$-based estimators exhibit a substantially larger scatter ($\sim$0.5 dex). For the least massive BH in our sample ($M_{\rm BH, RM} \sim 10^{7.5} M_\odot$), which is accreting at a rate close to the Eddington limit ($\lambda_{\rm Edd} = 0.8$), all SE calibrators return a BH mass that is an order of magnitude larger than the RM estimate. This discrepancy may indicate a systematic overestimation of BH masses for highly accreting BHs at high redshift.

Finally, using our new measurements together with other high-redshift sources with independently estimated BH masses, we provide the first high-redshift calibrations of SE BH mass estimators based on H$\alpha$ and H$\beta$. Although a larger and broader BH-mass sample is needed to further reduce the parameter uncertainties, our calibration can already be applied to the newly discovered JWST BH population in the early Universe.
}


   \maketitle
%
%-------------------------------------------------------------------

\section{Introduction}

The majority of massive galaxies, both at high redshifts and in the local Universe, are thought to host supermassive black holes (BHs) with masses exceeding $10^6-10^9 M_\odot$ at their centers \citep{Magorrian:1998}. The mass of the BH ($M_{\rm BH}$) correlates with several properties of the host galaxy (such as the stellar mass of the bulge, that of the entire galaxy, or the stellar velocity dispersion; \citealt{Kormendy:2013}), suggesting a common evolutionary scenario between the assembly of galaxies and the growth of their BHs.
Despite the widely accepted coevolution framework between the BH and galaxy growth \citep{Heckman:2014}, supported by the tight correlations observed in the local Universe, the validity of the $\rm M_{BH} - M_\star$ relation remains debated.
In particular, while the majority of local galaxies appear to lie on the relation, with a typical ratio M$_\star$/M$_{\rm BH}$$\sim$0.001 \citep[e.g.,][]{Kormendy:2013, Reines:2015}, some BHs are found to be overmassive with respect to the stellar mass of their hosts \citep[e.g.,][]{Bodgan:2012}.

At cosmic noon ($z\sim1-3$), several studies have reported the presence of overmassive BHs \citep{Trakhtenbrot:2015, Mezcua:2024}. This led to a contrasting picture, with some studies pointing to an evolution of the $\rm M_{BH} - M_\star$ relation with redshift \citep{Merloni:2010, Bennert:2011, Zhang:2023}, implying that BHs grow more quickly than their hosts at early cosmic times, while other works claim that such an evolution did not occur until $z\sim 4$ \citep{Shen:2015,Suh:2020, Sun:2025_z4}.

The bright z$\sim$6 quasars (QSOs) are typically overmassive with respect to their hosts \citep[e.g.,][]{Wang:2013, Pensabene:2020}; however, this is believed to be due to a selection bias toward the most luminous BHs \citep{Lauer:2007}.
Indeed, once this selection bias is taken into account, the tension with the local relation decreases
\citep{Schulze:2014}.

The \textit{James Webb} Space Telescope (JWST; \citealt{Gardner:2023}) has revealed large populations of overmassive BHs at high redshift ($z>4$),  not limited to bright QSOs.
In particular, its unprecedented sensitivity has enabled the discovery of a large number of faint ($L_{\rm bol} \sim 10^{43-45}$ erg~s$^{-1}$) broad-line active galactic nuclei (BLAGN)  at $z>4$ that appear to deviate from the local scaling relations, being overmassive with respect to their hosts \citep{Harikane:2023,Matthee:2023,Ubler:2023, Maiolino:2024, Juodzbalis:2026_jadessample}.
Some targets show extreme stellar-to-BH mass ratios approaching unity \citep{Kokorev:2023,Juodzbalis:2024}.
This new population of BLAGN comprises $\sim$ 30$\%$ of little red dots \citep[LRDs;][]{Matthee:2023}, while the remainder are little blue dots \citep[LBDs;][]{Brazzini:2026}, sharing similar sets of properties (compact hosts, broad permitted line, X-ray non-detections, and overmassive BHs), but different spectral energy distributions \citep[see][ for more details]{Brazzini:2026}.
However, recent works suggest that LRDs and LBDs may belong to the same population, observed at different inclinations \citep{Madau:2026}.

Unlike high-redshift QSOs, even when accounting for observational biases, these BHs remain overmassive with respect to local relations \citep[e.g.,][]{Juodzbalis:2024}, implying an evolution of the $\rm M_{BH} - M_\star$ relation at $z>4$ \citep{Pacucci:2023, Jones:2025,Sun:2025}. However, other works argue that observational biases may still be underestimated and therefore no intrinsic evolution should be invoked to explain this population of BHs \citep{Li:2025, Ziparo:2026}. Stacking analyses also identify a population of BHs that are not overmassive \citep{Geris:2026}.

Regardless of whether the local scaling relations evolve or not, a large variety of evolutionary scenarios and models have been proposed to explain the presence of such massive BHs in the early Universe. 
If the newly discovered BLAGN are indeed overmassive, heavy BH seeds \citep{Natarajan:2024,Cenci:2025} or super-Eddington accretion rates \citep{Trinca:2024, King:2025} are required. Despite the puzzling nature of these sources, which still calls for a comprehensive physical description, their overmassive nature should be carefully tested.
In particular, the stellar masses of these BLAGN are highly uncertain and depend on the fraction of light attributed to stellar emission \citep[e.g.,][]{Wang:2024_stellarpop, Akins:2025, Leung:2025}. A further major concern is that the BH masses may be overestimated \citep[e.g.,][]{Lambrides:2024, Naidu:2025_BHstar} due to the questionable applicability of typical calibrations in this new class of objects. 
Specifically, the typical relations used to infer \Mbh are calibrated on local active galactic nuclei (AGN) and then extrapolated to the high-$z$ and high-luminosity regimes, where their applicability is still a matter of debate \citep[e.g., ][]{Bertemes:2025B, Bosman:2025}, and they feature typical uncertainties on the order of $\sim 0.4-0.5$ dex \citep{shen2013mass, Shen:2024}.
The discovery of this larger population of overmassive BHs in the early Universe by JWST has raised questions about the validity of these relations, highlighting the need for more accurate BH mass measurements at high redshifts.
Such measurements are crucial not only for understanding how galaxies and their BHs coevolve but also to shed light on the formation of supermassive BHs at high redshift, providing key insights into the nature of BH seeds and their initial masses.

For local BHs, the mass can be robustly measured through kinematic methods that require resolving the sphere of influence of the BH, such as exploiting stellar kinematics \citep{Gebhardt:2000, Nguyen:2025}. 
Alternatively, we can use the detection of emission lines arising from clouds in the broad-line region (BLR), which are photoionized by radiation from the accretion disk. Under the assumption of a virialized BLR, the BH mass can be estimated as

\begin{equation}
    M_{\rm BH} = \frac{fV^2R}{G}, 
\end{equation}

\noindent where V is the rotational velocity of the BLR clouds and R is the distance between the BLR and the ionizing source.
However, both the geometry of the BLR and the inclination of the line of sight are, in most cases, unconstrained. To account for this, the $f$ factor encapsulates the BLR geometry, kinematics, and inclination of the line of sight. 
The rotational velocity ($V$) can be estimated using the width of the broad permitted lines. Different line-width measurements have been explored
in the literature, such as the full width at half maximum (FWHM) or the line velocity dispersion (see e.g., \citealt{peterson2004central}),  each of which has practical strengths and weaknesses.

The FWHM of the broad emission lines can be measured directly in observations, while the radius of the BLR can be estimated using reverberation mapping (RM) campaigns \citep[see, e.g.,][and references therein]{Dallabonta:2020}. In particular, we can model the measured time lag between variations in continuum flux and variations in broad-line luminosity, assuming a given distribution of the BLR clouds, to derive the BLR radius.
The RM approach requires accurate and frequent measurements of the BLR and continuum luminosities over a span of months, which at high redshift can become decades in the observed frame due to cosmological time dilation. 
Hence, obtaining high-$z$ measurements of BLR sizes and RM-derived BH masses can be achieved through long-term monitoring campaigns \citep{Shen:2024, McDougall:2025}.
Although RM campaigns are resource-intensive, they also provide a relation between the BLR radius (R) and the monochromatic luminosity at some wavelengths (e.g., 5100 \AA), thus providing a robust proxy for the BLR radius \citep[$R-L$ relation; ][]{Kaspi:2000, Bentz:2009}.
This discovery has enabled the so-called single epoch (SE) virial BH mass calibrations, where a single spectrum can provide both quantities needed to compute $M_{\rm BH}$, namely the FWHM of the line and the continuum luminosity, respectively representing the rotational velocity ($V$) and the BLR radius (R).

Other attempts have been made to obtain independent BH mass measurements with the near-IR GRAVITY interferometer \citep{Abuter:2017}, which can spatially resolve the BLR and measure its kinematic properties. Modeling of BLR kinematics yields a dynamical BH mass \citep[e.g.,][]{Sturm:2018, GRAVITY:2020_iras091, GRAVITY:2024}. However, this technique is currently limited to only very luminous QSOs with a nearby star for adaptive-optics (AO) correction \citep{Gravity:2022}.

All the techniques mentioned above for obtaining a direct measure of the \Mbh have, in recent years, extended their reach to the high-$z$ Universe \citep{Abuter:2024, Shen:2024,Liao:2025, Gravity_z4:2025}, but the $z>4$ Universe has so far remained inaccessible with these methods, except for a dynamical BH mass measurement at $z\sim7$ in a lensed LRD \citep{Juodzbalis_2025_dynamical}.
Thus, to derive the BH masses for statistical samples at $z>4$, we can only rely on SE virial BH mass estimators.
With this approach, \Mbh is derived using a single spectroscopic measurement, which yields the width of a broad line and the continuum (or line) luminosity.

In this work, our aim was to test the reliability of SE BH mass estimators in the high-$z$ Universe by directly comparing them with robust RM-based measurements. We focus on a sample of five QSOs at z$\sim$2 with RM BH masses \citep{Shen:2024}, and we take advantage of the unprecedented sensitivity and spatial resolution of JWST/NIRSpec Integral Field Unit (IFU) data to obtain high-quality measurements of both the \hb and \ha broad emission lines. This allows us to perform a direct comparison between two different SE estimators and assess their consistency at high redshift. 
To achieve this goal, we implement a new data-reduction strategy for the medium-resolution gratings that effectively extends the usable wavelength coverage to longer wavelengths. 
With this method, we recover the \ha\ emission line using archival data whose spectroscopic coverage, based on standard JWST data, was limited to \hb and \oiii\ emission.

The paper is structured as follows. In Sect. \ref{sec:datared}, we present the methods used for the new data reduction.  In Sect. \ref{sec:data}, we describe the analysis of the sample of five QSOs. In Sect. \ref{sec:comparing_se_masse}, we estimate the BH masses using some of the most widely used SE estimators. In \ref{sec:calibrating_se} we derive our own relation calibrated at $z\sim2$.
We draw our conclusion in Sect. \ref{sec:conclusions}.
Throughout this work, we adopt the cosmological parameters from \citet{Planck:2015}: $H_0
= 67.7$ \kms\,Mpc$^{-1}$, $\Omega_{\rm m}$ = 0.307, and $\Omega_\Lambda$= 0.691.

%--------------------------------------------------------------------
\section{Changes to the standard data reduction}
\label{sec:datared}

\begin{figure*}
    \centering
    \includegraphics[width=0.85\linewidth]{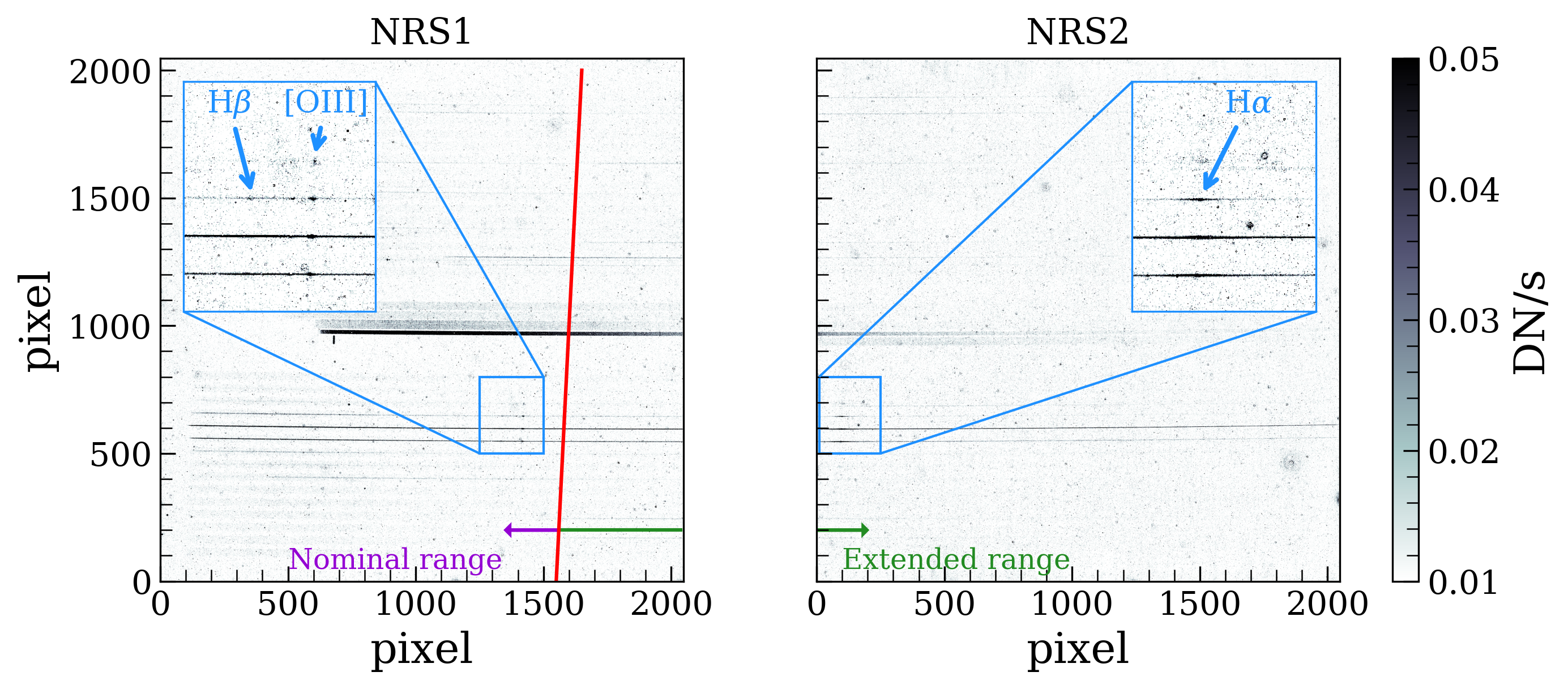}
    \caption{Count-rate images for the G140M/F100LP observations of RM332, a QSO at $z\sim 2.6$ downloaded from MAST, for the NRS1 (left) and NRS2 (right) detectors.
    The vertical red line indicates the maximum wavelength range of the nominal observational coverage.
    The inset panels show zoomed-in views of the \oiii-\hb complex on the left and \ha emission on the right.
    }
    \label{fig:count_rate_maps}
\end{figure*}

Data from NIRSpec Intregral Field Spectroscopy (IFS) medium-resolution observations (G140M/F070LP, G140M/F100LP, G235M/F170LP, G395M/F290LP), with resolving power $\rm{R}=\lambda/\Delta\lambda$ varying between $\sim 700$ and $\sim 1300$ (average R$\sim$1000; \citealt{Jakobsen:2022}) occupy only the first detector (NRS1), while the second detector (NRS2) is not used in the standard reduction process because it contains spectra beyond the nominal wavelength range.
However, we detect emission lines in the second detector and continuum emission up to the red end of NRS2 (Fig. \ref{fig:count_rate_maps}). The figure shows count-rate maps downloaded from the Mikulski Archive for
Space Telescopes (MAST) for the G140M/F100LP observations from PID: 2057, targeting a QSO at $z\sim 2.6$. 
The count-rate maps reveal the presence of the \oiii\ and \hb emission lines in the nominal wavelength range, but also  \ha\ emission in the second detector beyond the nominal wavelength range of G140M/F100LP observations.

This work extends the wavelength range of medium-resolution gratings to recover the emission in the second detector. Extending to the second detector effectively provides the equivalent of two grating/filter observations in a single observation. For the G140M/F100LP grating/filter combination, we extend the nominal wavelength range of 0.97--1.88 $\mu$m to 3.55 $\mu$m, corresponding to an increase of 1.67 $\mu$m in spectral coverage. Similarly, for the G235M/F170LP we extended the spectra up to the maximum NIRSpec wavelength of 5.27$\mu$m, resulting in an increase of $\sim$2.16$\mu$m, which corresponds to a 150\% increase in effective spectral coverage. We report the nominal and extended range for all filters in Table \ref{tab:new_coverage}.

Previous studies demonstrate spectral extension for the high-resolution (R$\sim$2700) grating \citep{Deugenio:2025_QSO1, Torralba:2025}, but the resulting wavelength gain is much smaller, on the order of 0.1 $\mu$m.
The extension is made possible by the fact that the photon conversion efficiency does not drop dramatically as a function of wavelength, particularly for the G140M/F100LP and G235M/F170LP grating/filter configurations \citep[for details see][]{Boker:2022}, which benefit most from the extension.
For the G395M/F290LP, extending the wavelength range beyond the nominal limit of 5.27$\mu$m is feasible \citep{Pascalau:2026}; however, it is limited by the detector sensitivity, which drops significantly, reaching zero at  $\lambda\sim5.5\mu$m \citep{Jakobsen:2022}.
The G140M/F070LP could, in principle, be extended; however, it offers limited scientific gain and requires significantly more effort.
For IFU observations, the wavelength range coverage of G140M/F070LP is limited to $0.90-1.26$ $\mu$m, because the spectrum falls off NRS1 at bluer wavelengths. Hence, the only effective gain of using G140M/F070LP with respect to G140M/F100LP is to observe in the $0.90 - 0.97$ $\mu$m interval. Additionally, using F070LP introduces second- and third-order spectral contamination at shorter wavelengths, requiring correction for fourth- and fifth-order contamination. For these reasons, we do not consider this setup in this work.

In this section, we describe the modification to the standard JWST data reduction pipeline and reference files that allowed us to extend the wavelength range covered by NIRSpec/IFU R1000 observations, together with all the tests performed to ensure the validity of the modified pipeline.
Finally, we provide an estimate of the RMS and the performance achievable with these observations.

\subsection{Data reduction}

To extend the medium-resolution gratings, we first created and modified the reference files to extract flux beyond the nominally calibrated wavelength range and then performed the flux calibration.
For the initial tests and calibration steps, we used observations of standard stars from calibration programs 6645, 1536, 1537, and 1538. The four programs observed three different stars with all the possible configurations of gratings and filters, with the goal of calibrating the flux and the flat fields for the IFU and other NIRSpec modes. We also used observations of other targets (local galaxies, AGN, and high-$z$ galaxies) obtained with different gratings to verify the accuracy of the calibrations and to calibrate the flux of our extension of the medium-resolution gratings, which we then applied to other targets.
We selected the complementary sources used to verify and calibrate our extension to respect the following criteria: (i) they have observations with at least two consecutive medium-resolution gratings (i.e., G140M/F100LP + G235M/F170LP or G235M/F170LP + G395M/F290LP); (ii) they are bright enough to have continuum detection with $S/N>5$ across the full wavelength range covered.

For all our reductions, we retrieved the count-rate maps (from both the first and second detectors) from the MAST archive. We then applied the standard data-reduction steps \verb|calwebb_spec2| and \verb|calwebb_spec3| using pipeline version 1.18.1 and the context \verb|jwst_1364.pmap|, performing an initial reduction with the standard pipeline as a benchmark. 
We then modified the reference files to allow for the extension.
Specifically:

\begin{itemize}
    \item We extrapolated each \verb|S-flat| above the nominal wavelength range to unity (see Sect. \ref{sec:sflat}).    \item We modified the \verb|F-flat| files so that each file is extended up to 5.5$\mu$m by concatenating the \verb|F-flat| files for all filters (see Sect. \ref{sec:fflat}).
    \item We modified the wavelength range for each grating in the \verb|nirspec_cube_par| and  \verb|nirspec_wavelength_range| reference files. 
\end{itemize}

In the following subsections, we provide a detailed description of the main changes made to the various configuration files.
Moreover, the pipeline itself required an additional modification: we removed the check for the presence of files from the second detector in the ``\verb|calwebb_spec2|'' step of the pipeline, which otherwise causes an error and stops the data reduction. 
All modified reference files, together with a guide on how to apply the changes to the official pipeline and the Python scripts to perform the flux calibration, are available on GitHub\footnote{\href{https://github.com/eleonoraparlanti/nirspecIFU-extended}{https://github.com/eleonoraparlanti/nirspecIFU-extended}}.

With all these modifications, the pipeline successfully creates the cubes up to the maximum wavelength range covered by the second detector. The extension naturally creates a ``gap'' in the final spectra, similar to that seen in data obtained with the high-resolution gratings, due to the physical gap between the two detectors.
Table \ref{tab:new_coverage} reports the extended and nominal wavelength coverage, as well as the wavelength range that falls into the detector gap in the extended data.

\renewcommand{\arraystretch}{1.4}  
\begin{table}[ht!]
\footnotesize
    \centering
        \caption{Nominal and extended science ranges, and new gap, for each filter and grating combination analyzed in this work.}
\begin{tabular}{c|c|c|c}
 \hline  \hline
         Grating/Filter&Nominal range& Extended range&Gap\\
 & [$\mu$m]& [$\mu$m]&[$\mu$m] \\  \hline
         G140M/F100LP& 
      0.97 -- 1.88& 0.97 -- 3.55 & 2.17 -- 2.28 \\ 
 G235M/F170LP & 1.66 -- 3.15& 1.66 -- 5.27& 3.66 -- 3.82\\
  \hline
\end{tabular}

    \tablefoot{Nominal range from \citet{Boker:2022}.}
    \label{tab:new_coverage}
\end{table}

\subsubsection{Spectroscopic flat (S-flat)}
\label{sec:sflat}
The \verb|S-flat| files consist of one file for each detector and each filter/grating configuration and account for all the losses in the optical path from the aperture plane to the disperser.
Each file consists of three image extensions and two binary tables. 
The small corrections, slowly varying with wavelength, are stored in the SCI extension, while the binary table extension labeled ``FAST\_VARIATION'' encapsulates rapidly wavelength-dependent variations with larger relative amplitudes.

To extend the wavelength range of the reduced data, we modified the SCI and DQ extensions, which contain the detector response at each pixel and the data-quality flags.
In the IFU, for each grating and filter, the various wavelengths fall on the same detector pixel; hence, pixels containing wavelengths outside the nominal range have a value of ``nan'' in the SCI flat and are flagged as pixels not to be used in the DQ extension.
For the S-flat in NRS1, we extended each trace until the end of the detector in the SCI and DQ images. 
Because the spectral traces exhibit slight curvature and tilt on the detector, we fitted a linear polynomial to their edges and extended the resulting paths according to the best-fit model.
We fixed the pixels in the extended SCI version to unity, while in DQ, we removed each flag by setting the value to zero. 
We show the original S-flat for NRS1 and the extended version in Fig. \ref{fig:sflat_image}.
Moreover, the FAST\_VARIATION table, which is only defined over the nominal wavelength range, required modification. We extended it with a constant value from the last defined element to the maximum wavelength covered by the extended detectors. Figure~\ref{fig:sflat_image} (right panel)  shows the original and extended fast-variation values.

We retrieved the latest NRS2 S-flat files for each grating/filter combination from the  Calibration References Data System (CRDS) website. Since NRS2 was not used in the reduction of the medium-resolution IFU products, we set its DQ extension to DO NOT USE while we initialized the SCI extension to unity.
In this case, we only modified the DQ extension by removing the flag for each pixel, and we extended the FAST\_VARIATION table as described above.

\subsubsection{Fore-optics flat (F-flat)}
\label{sec:fflat}

The F-flat accounts for losses from all the reflections in the telescope and the NIRSpec fore optics. The F-flat does not depend on the detector but is defined for each filter and grating combination.
To extend the wavelength coverage, we concatenated the binary table extension describing the FAST\_VARIATION of one filter and grating combination with that of the following combination. For example, the newly defined F-flat for G140M/F100LP is now the concatenation of those of G235M/F170LP up to 3.15$\mu$m and with that of G395M/F290LP up to 5.2$\mu$m.

\subsection{Flux calibration}

With the aforementioned changes to the standard data reduction workflow, the output of the \verb|calwebb_spec3| step is a datacube with an extended wavelength range (see Table~\ref{tab:new_coverage}), although only the spectra within the nominal range are flux calibrated. We therefore determined an empirical correction curve that can be applied to the final datacube to calibrate the flux in the extended wavelength region. 

\begin{figure}[ht!]
    \centering
    \includegraphics[width=0.98\linewidth]{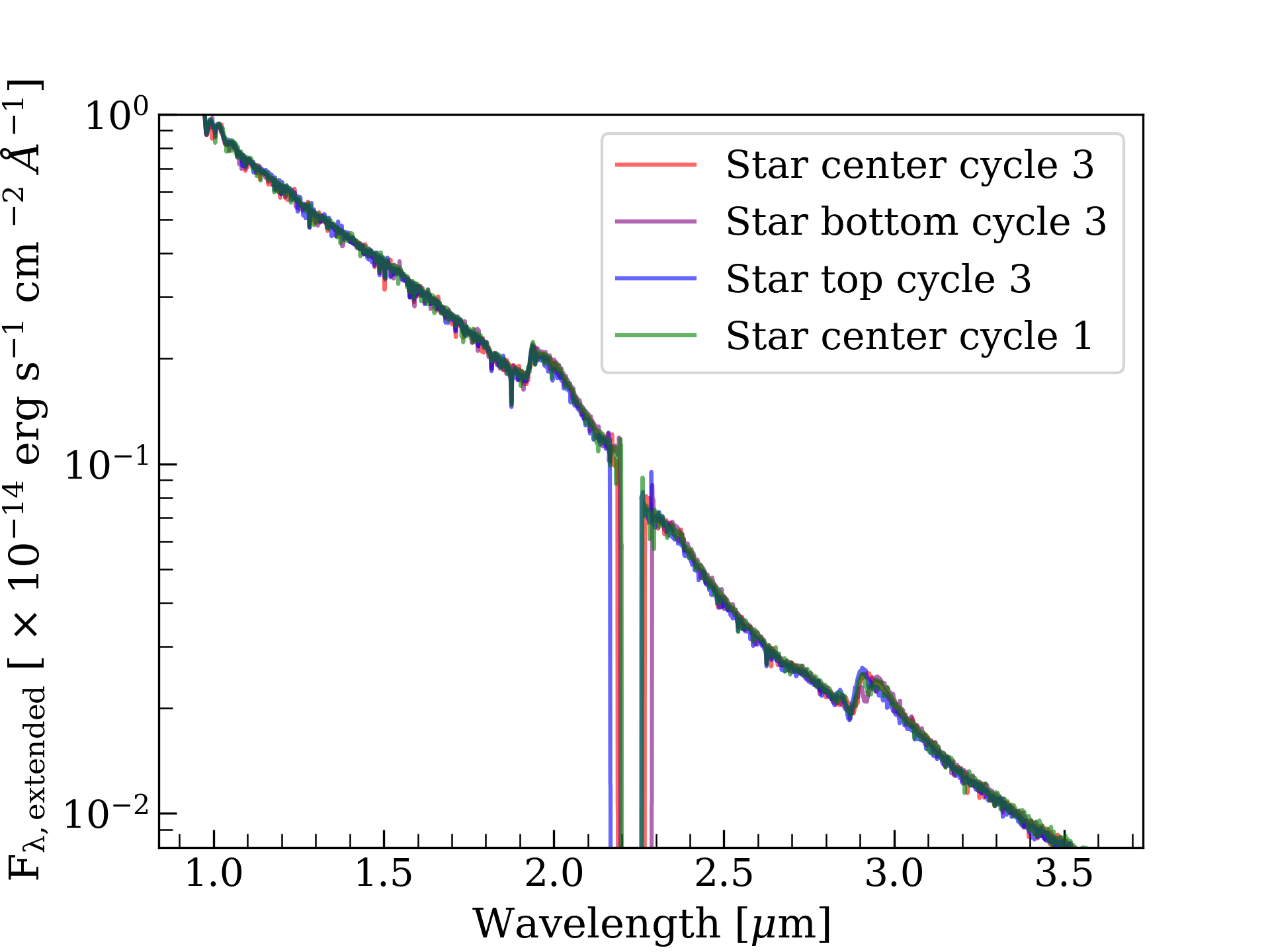}
    \caption{Observed flux for the star P330-E obtained with the extended G140M/F100LP filter, as part of programs 1538 and 6645.
    The different color curves show the same flux for the star observed in cycle 1 (green), in cycle 3 at the center of the FOV (black), and in the bottom (blue) and top (purple) parts of the FOV.
    The equivalent data for G235M/F170LP are shown in Appendix \ref{appendix:fluxcalibration235m}.
    }
    \label{fig:std_star_different_epoch}
\end{figure}

First, we verified that the profile of the uncalibrated spectra in the extended wavelength region does not depend on the location of the targets within the field of view. This verification ensures that the empirical correction curve we derive depends only on wavelength, with no spatial dependence. We thus used the data for the standard star P330-E, which was observed in cycles 1 and 3 across all grating and filter configurations in PIDs 1538 and 6645 (see also Table \ref{tab:claibration_target} for details of the dataset used for the calibrations). In PID 6645, the star was also observed by centering the star at different locations within the field of view.
Figure~\ref{fig:std_star_different_epoch} (\ref{fig:std_star_different_epoch_g235m}) shows the G140M/F100LP (G235M/F170LP) datacubes produced by \verb|calwebb_spec3|, with spectra extracted from a circular aperture of 0.3\arcsec\ centered on the star. Comparing the spectra, we find no significant differences for the same target observed two years apart and at different positions within the field of view. This result indicates that our empirical correction curve depends only on wavelength and can be applied to any spectrum extracted from any location within the field of view.

 We then estimated the empirical correction curve using spectra of various standard stars observed with all filter–grating configurations from calibration programs PID 6645, 1536, 1537, and 1538, as well as bright sources (QSOs and local galaxies) observed in programs 2564 and 2186. Table \ref{tab:claibration_target} details the programs and targets used for flux calibrations. We derived the correction curve by comparing the uncalibrated spectra in the extended wavelength region with the flux-calibrated spectra obtained from the filter–grating configurations covering longer wavelengths. To this end, we first extracted the flux-calibrated spectrum (hereafter referred to as the intrinsic spectrum) in each nominal wavelength range.
However, we find that dividing the intrinsic spectrum by the uncalibrated spectrum is not sufficient to obtain an accurate correction curve. In particular, contamination from higher–order spectra (primarily second and third order) must be accounted for, as its contribution becomes significant at wavelengths around $n,\lambda_{\rm min}$, where $n$ is the spectral order and $\lambda_{\rm min}$ is the shortest wavelength of the filter. 

Assuming contributions only up to the third order, the output of \verb|calwebb_spec3| can be expressed as

$$
\begin{cases}
S_\lambda(\lambda) = f_\lambda(\lambda) \qquad \lambda\le\lambda_{\rm max}^{\rm nominal} \\
S_\lambda(\lambda) \approx  k(\lambda)f_\lambda(\lambda)+ \alpha(\lambda)g_\lambda(\lambda)+\beta(\lambda)h_\lambda(\lambda) \quad \lambda>\lambda_{\rm max}^{\rm nominal},
\end{cases}
$$
where $\lambda_{\rm max}^{\rm nominal}$ denotes the maximum wavelength of the NIRSpec nominal range, $f_\lambda(\lambda)$ represents the intrinsic source spectrum, and $g_\lambda(\lambda)$ and $h_\lambda(\lambda)$ correspond to the second- and third-order spectra, respectively. The functions $k(\lambda)$, $\alpha(\lambda)$, and $\beta(\lambda)$ describe the wavelength-dependent transmission efficiencies for each spectral order. The second- and third-order spectra can also be expressed as functions of the first-order spectrum, such that $g_\lambda(\lambda) \propto f_\lambda(\lambda/2)$ and $h_\lambda(\lambda) \propto f_\lambda(\lambda/3)$. Therefore, the uncalibrated spectrum beyond the nominal wavelength range can be written as
$$
S_\lambda(\lambda) = k(\lambda)f_\lambda(\lambda)+ \tilde\alpha(\lambda)f_\lambda(\lambda/2)+\tilde\beta(\lambda)f_\lambda(\lambda/3),
$$
where $\tilde{\alpha}(\lambda)$ and $\tilde{\beta}(\lambda)$ are the effective transmission coefficients for the second- and third-order contributions as function of $f_\lambda$. Dividing both sides by the intrinsic spectrum $f_\lambda(\lambda)$ gives
\begin{align*}
R(\lambda) = \frac{S_\lambda(\lambda)}{f_\lambda(\lambda)}&= k(\lambda)+ \tilde\alpha(\lambda)\frac{f_\lambda(\lambda/2)}{f_\lambda(\lambda)}+\tilde\beta(\lambda)\frac{f_\lambda(\lambda/3)}{f_\lambda(\lambda)}\\
 &= k(\lambda)+ \tilde\alpha(\lambda)r_2(\lambda)+\tilde\beta(\lambda)r_3(\lambda),
\end{align*}
where $r_2(\lambda)$ and $r_3(\lambda)$ are known functions determined by the intrinsic source spectrum $f_\lambda(\lambda)$. At a fixed wavelength $\bar{\lambda}$, there are three unknown variables: $k(\bar{\lambda})$, $\tilde{\alpha}(\bar{\lambda})$, and $\tilde{\beta}(\bar{\lambda})$. Therefore, if both $S_\lambda(\lambda)$ and $f_\lambda(\lambda)$ are known for at least three independent sources, we can solve a system of equations at each wavelength for 
 $k(\lambda)$, $\tilde{\alpha}(\lambda)$, and $\tilde{\beta}(\lambda)$.

\begin{figure*}[ht!]
    \centering
    \includegraphics[width=0.82\linewidth]{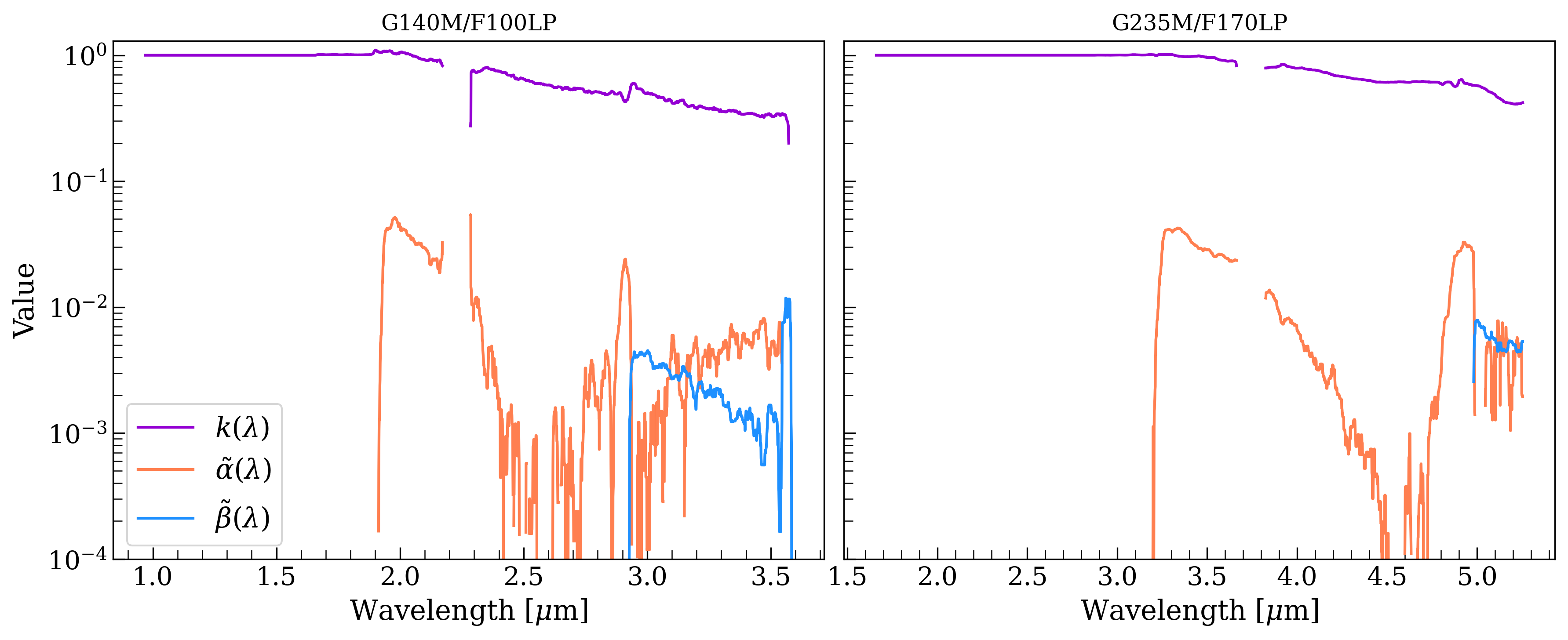}

    \caption{Values of $\tilde{\alpha(\lambda)}$ (orange), $\tilde{\beta(\lambda)}$ (blue), and $k(\lambda)$ (violet), as a function of wavelength for G140M/F100LP (left) and G235M/F170LP (right).
    }
    \label{fig:calibrators}
\end{figure*}

We recovered the functions  $k(\lambda)$, $\tilde{\alpha(\lambda)}$, and $\tilde{\beta(\lambda)}$ for G140M/F100LP and G235M/F170LP by using all datasets for which we have both the extended spectrum $S_\lambda(\lambda)$ from our reduction and the intrinsic spectrum $f_\lambda(\lambda)$, obtained by reducing the three gratings within their nominal wavelength range. 
For each combination of three datasets, we computed the three functions. To account for bad pixels and spikes in these calibration functions, we averaged and smoothed them over a window of 40 channels along the wavelength axis. We report the three functions in Fig. \ref{fig:calibrators}.
We note that the $\tilde{\alpha}$ and $\tilde{\beta}$ functions, which quantify the level of the contamination from the second- and third-order spectra in the observed spectrum, respectively, are on the order of 1--5\% at their maximum, occurring in the range between 2--2.3 $\mu$m for G140M and 3.2 -- 4 $\mu$m for G235M.

Finally, using the calibration functions, we obtained the intrinsic spectrum of each source in the extended region as

\begin{equation}
\label{eq:spectrum_final}
    f_\lambda(\lambda) = \frac{[S_\lambda(\lambda) - \Tilde{\alpha}(\lambda)f_\lambda(\lambda/2)
    - \Tilde{\beta}(\lambda)f_\lambda(\lambda/3)] } {k(\lambda)},
\end{equation}

\noindent where $\Tilde{\alpha}(\lambda)$ and $\Tilde{\beta}(\lambda)$ are defined as zero at wavelengths smaller than $2\lambda_{\rm min}$ and $3\lambda_{\rm min}$, respectively; and $k(\lambda)$ is defined as unity at $\lambda\le\lambda_{\rm max}^{\rm nominal}$.

\begin{figure*}[ht!]
    \centering
    \includegraphics[width=0.82\linewidth]{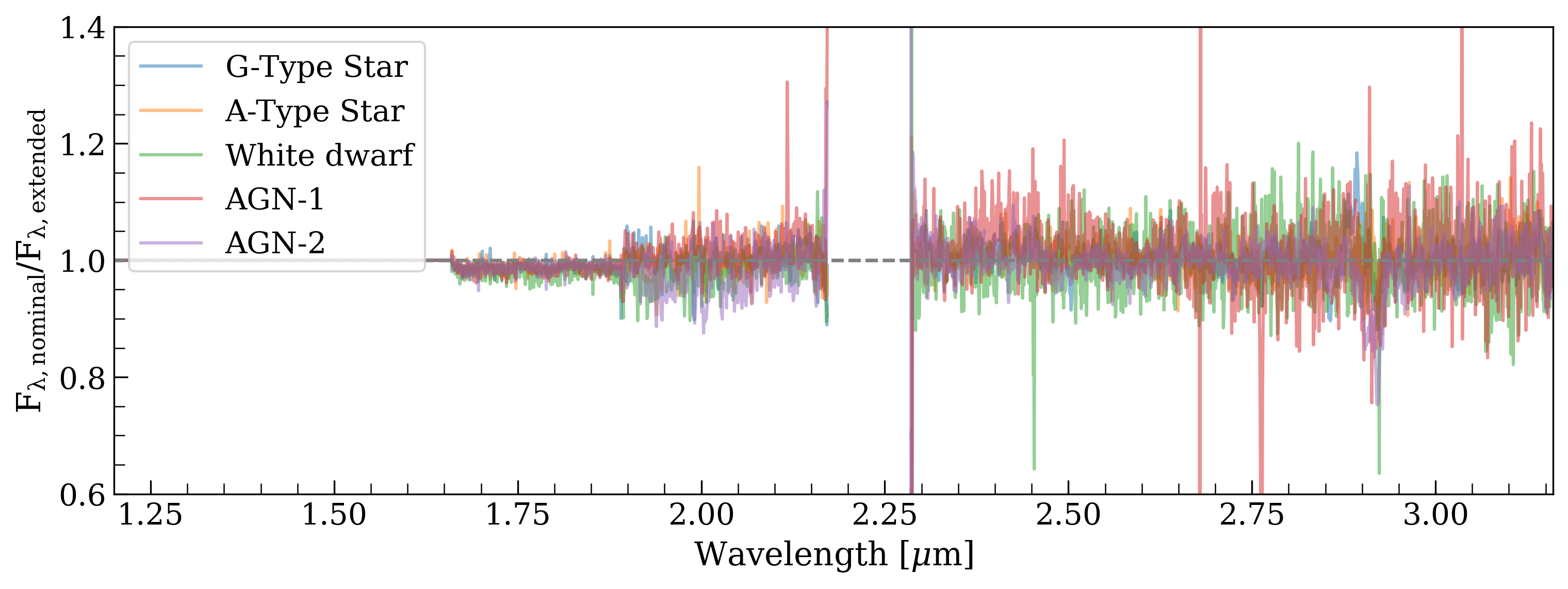}
    \caption{Ratio between the observed flux in the nominal wavelength range of the filters and the flux observed with the extended G140M/F100LP configuration. Figure \ref{fig:second_order_correction_g235m} shows the same ratio for G235M/F170LP.
    }
    \label{fig:second_order_correction}
\end{figure*}

To assess the accuracy of Eq.~\ref{eq:spectrum_final}, and therefore the reliability of the flux calibration in the extended wavelength range, Fig.~\ref{fig:second_order_correction} shows the ratio between the spectra obtained in the nominal ranges and the extended flux-calibrated spectra for G140M/F100LP, while Fig.~\ref{fig:second_order_correction_g235m} shows the same for G235M/F170LP. 
To perform this test, we used the data from the standard stars and two AGN, SDSSJ0749 and SDSSJ0841, which we labeled AGN 1 and AGN 2 from program 2654 (see \ref{tab:claibration_target} for details).
Figures \ref{fig:second_order_correction}  and  \ref{fig:second_order_correction_g235m} show that the calibration described above achieves an accuracy of $\sim 5\%$ and $\sim 10\%$, respectively, for the G235M/F170LP and G140M/F100LP configurations.
These values are well within the current NIRSpec flux calibration uncertainties and discrepancies between different gratings \citep[on the order of $\sim 20-25\%$;][]{deugenio:2025_dr4,Scholtz_dr4:2025}.

\subsection{Properties of the extended spectra}

\begin{figure*}[ht!]
    \centering
    \includegraphics[width=0.46\linewidth]{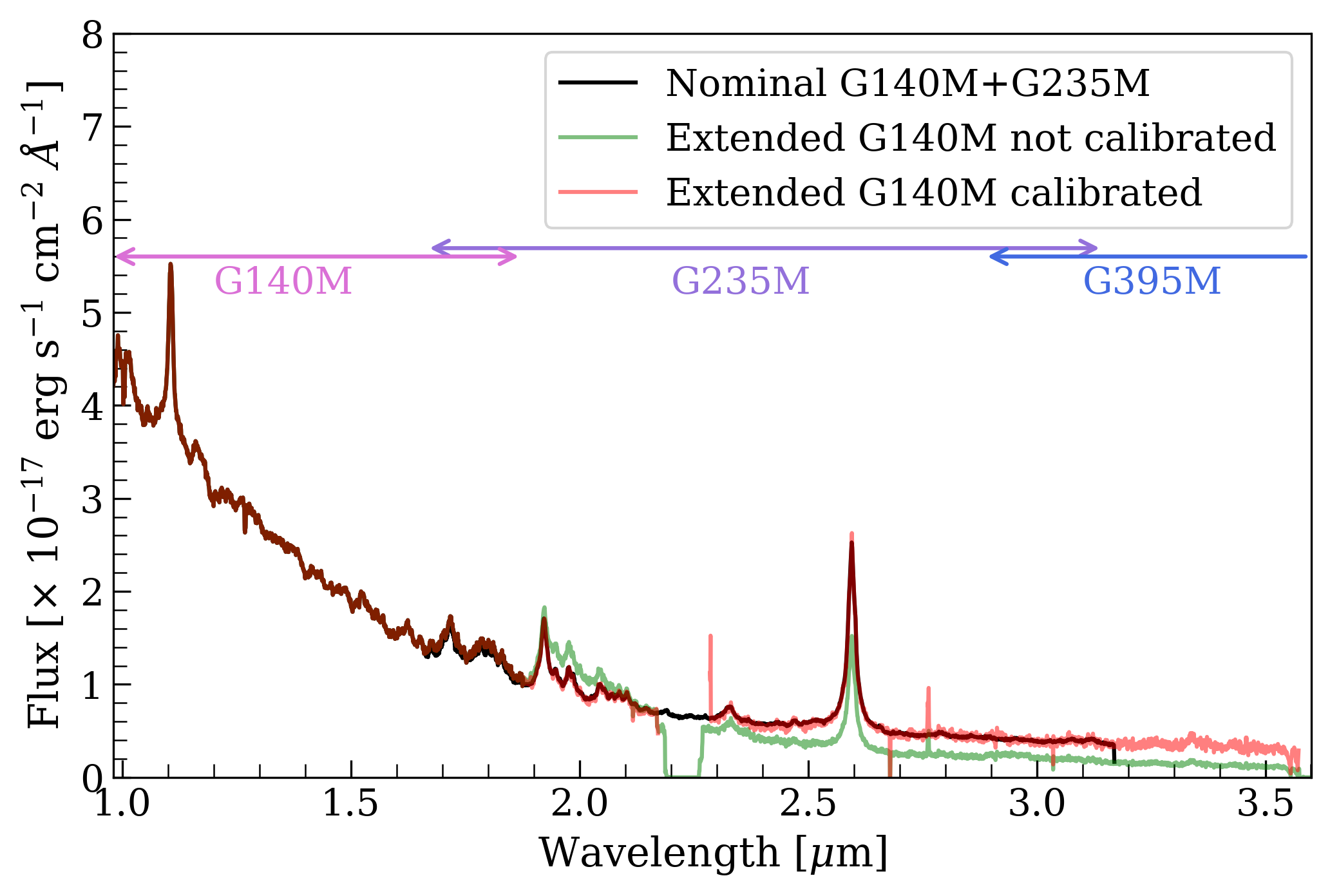}
    \includegraphics[width=0.48\linewidth]{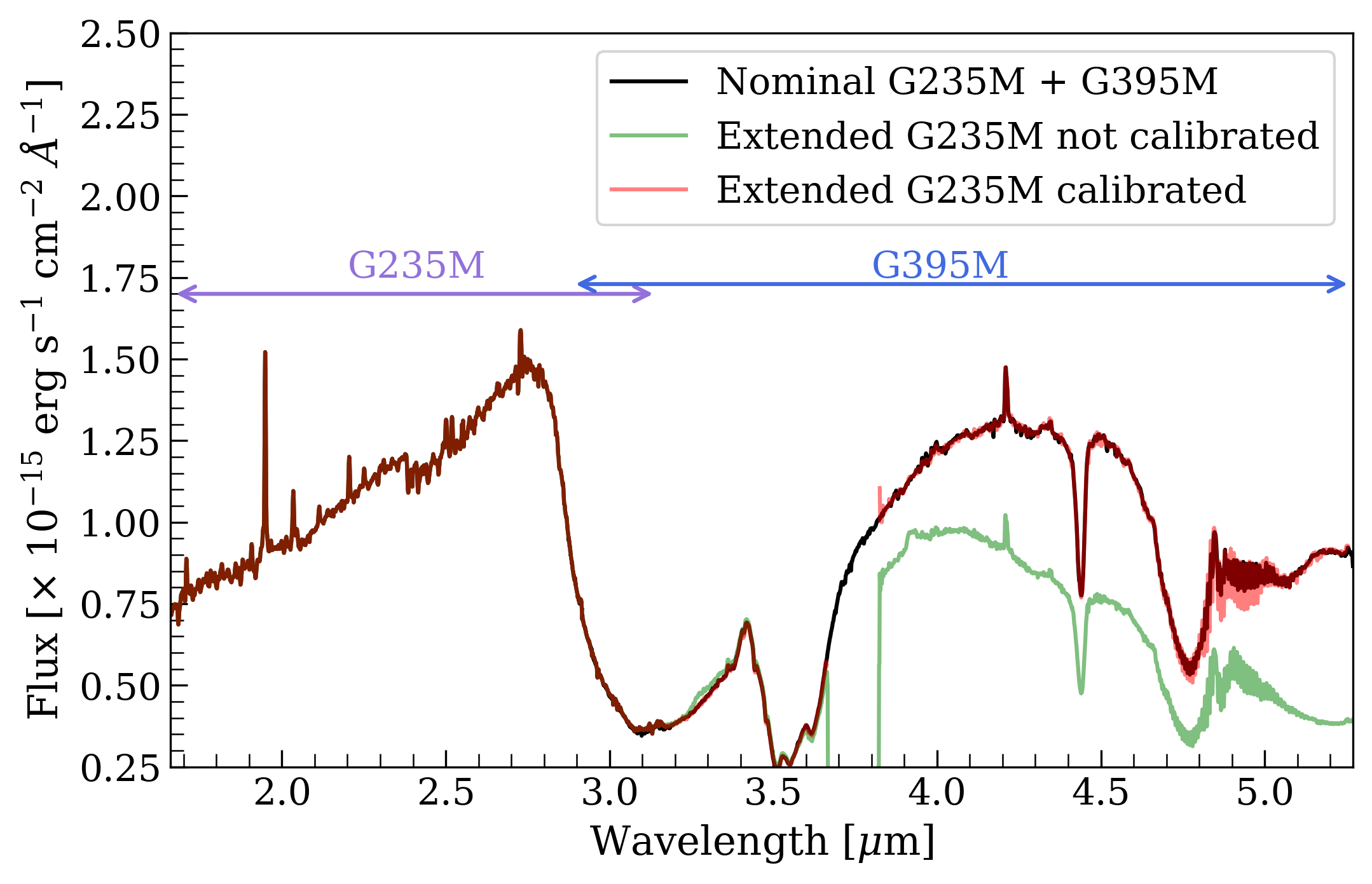}
    \caption{
    Examples showing the spectrum in the nominal range (black) and the extended spectra before and after the calibration reported in Eq. \ref{eq:spectrum_final} (green and red, respectively).
    Top: Spectrum extracted from a circular aperture of radius 0.5$^{\prime\prime}$ centered on the brightest pixel of AGN SDSSJ0841 (PID: 2654), the nominal G140M/F100LP+G235M/F170LP spectrum, and the extensions of the G140M/F100LP.
   Bottom panel: Spectrum extracted from a circular aperture of radius 0.5$^{\prime\prime}$ centered on the brightest pixel of ULRIG UGC-5101 (PID: 2186) including the nominal G235M/F170LP and G395M/F290LP spectra (black) and the extended G235M/F170LP spectrum.
    The nominal wavelength range of each filter is shown by the pink, purple, and blue lines for G140M/F100LP, G235M/F170LP, and  G395M/F290LP, respectively
    }
    \label{fig:example_spectra}
\end{figure*}

After successfully creating the cubes and computing the functions necessary to account for the second- and third-order spectra, and correctly calibrating the flux, we obtained the datacube containing only the target spectrum in the extended wavelength range.
Figure \ref{fig:example_spectra} shows two examples of the extension for both G140M/F100LP and G235M/F170LP in the top and bottom panels, respectively.
The two panels show that, after calibration, we achieve good agreement with the nominal spectrum, consistently removing the effect of the second- and third-order spectra and correcting for the unknown flat-field level in the extended region of the first detector and the entire second detector.

These spectra show that the RMS of the extended region is higher than that of the nominal range. We note that for AGN SDSSJ0841, we used the same exposure time for both filters, while for the target UGC-5101 in the bottom panel, the exposure time for the observations with the G235M is double that of the G395M.
The increase in RMS is due both to the decrease in disperser transmission outside its nominal wavelength range (see JWST/NIRSpec documentation\footnote{\href{https://jwst-docs.stsci.edu/jwst-near-infrared-spectrograph/nirspec-instrumentation/nirspec-dispersers-and-filters}{https://jwst-docs.stsci.edu/jwst-near-infrared-spectrograph/nirspec-instrumentation/nirspec-dispersers-and-filters}}) and to the increase in spectral resolution. 
In particular, for G140M, the transmission drops from $0.9$ at $\sim 1\mu$m, to $\sim 0.2$ at $3.5\mu$m, the maximum wavelength covered in our extension.
Similarly, the transmission curve of the G235M disperser peaks at $0.9$ at $\sim2.1 \mu$m and drops to 0.3 at $5\mu$m.

\begin{figure*}[ht!]
    \centering
    \includegraphics[width=0.45\linewidth]{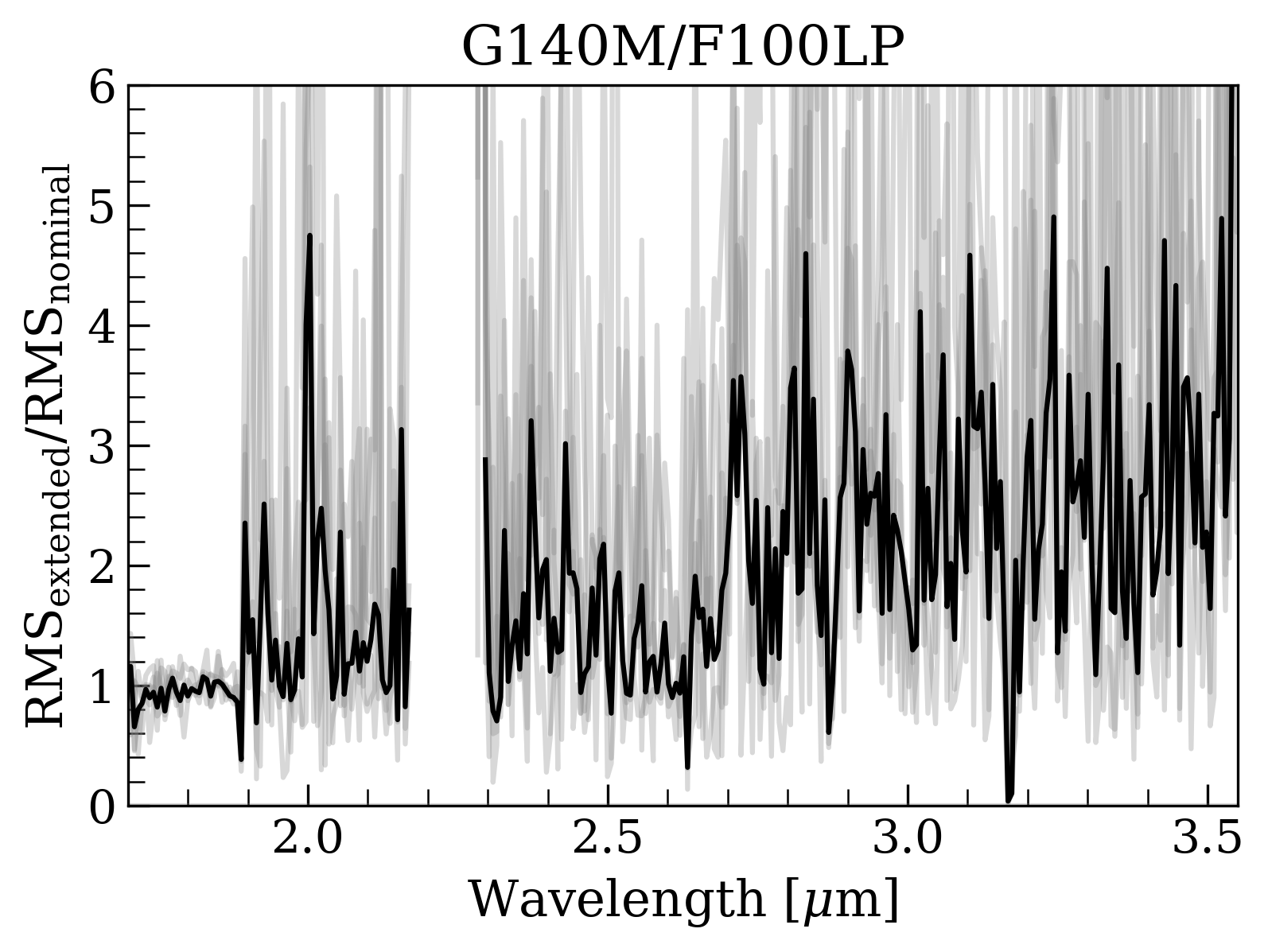}
    \includegraphics[width=0.45\linewidth]{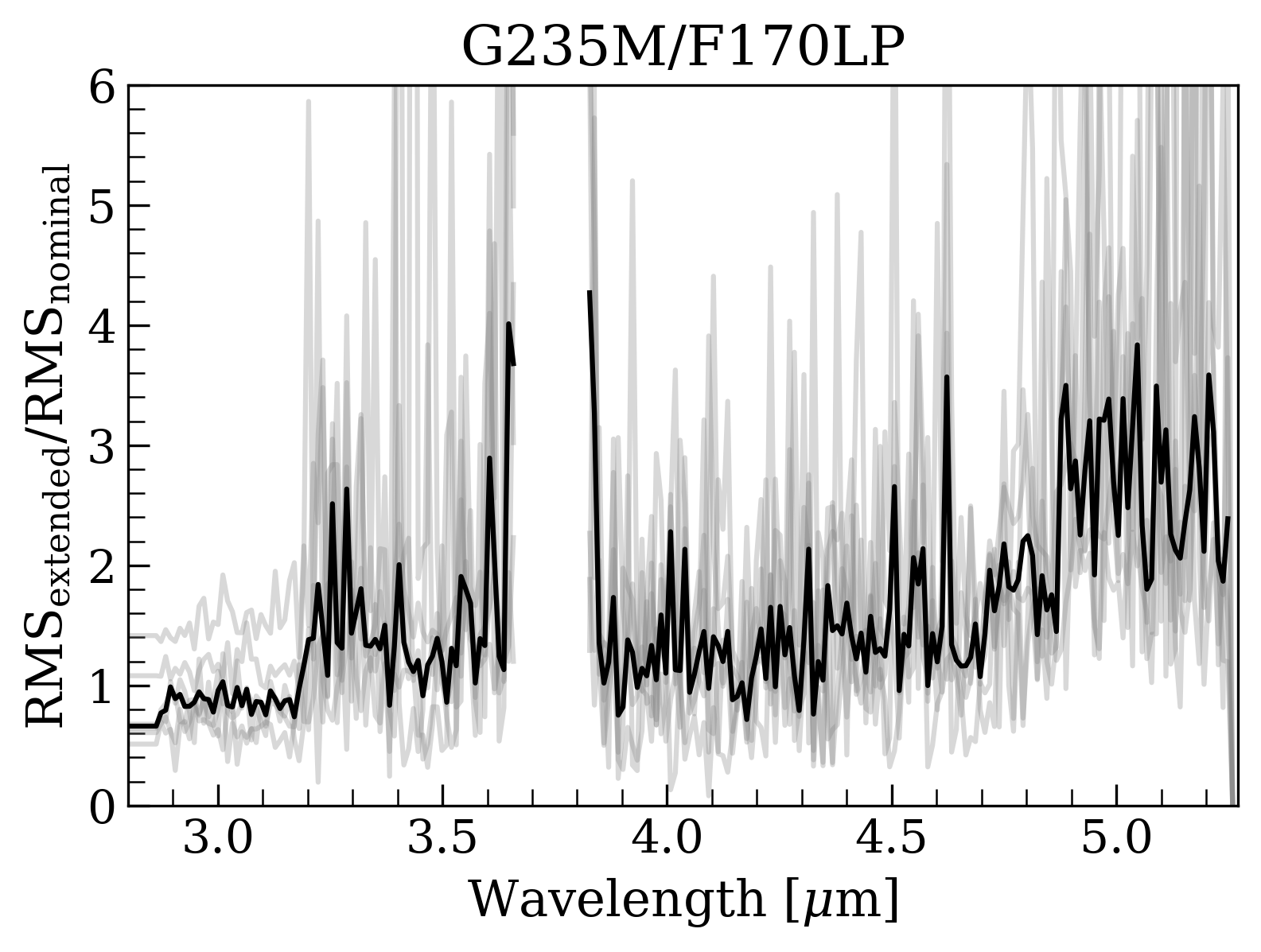}

    \caption{Increase in the RMS computed over 15 channels for the extended and nominal ranges of G140M/F100LP (left) and the G235M/F170LP (right).
    The solid line shows the mean ratio between the RMS in the extended version and that in the nominal range. 
    Gray lines show the RMS for each dataset with observations of all the filters and grating combinations at the same exposure time.
    }
    \label{fig:RMS}
\end{figure*}

We quantified the increase in the RMS with wavelength using the datasets with observations in all the grating and filter combinations, re-normalizing the measured RMS to account for differences in exposure time between the different ten wavelength channels for both the nominal and the extended spectrum.
Figure \ref{fig:RMS} shows the median ratio between the measured RMS of the extended cubes and that measured in the nominal range computed for each cube (shown in gray) as a function of wavelength, as a solid black line, for both G140M/F100LP and G235M/F170LP datasets. The value obtained for each dataset fluctuates, but the median steadily increases from one in the overlapping region between the two consecutive gratings to three at the redder wavelengths of our extension.

\begin{figure}[]
    \centering
    \includegraphics[width=0.98\linewidth]{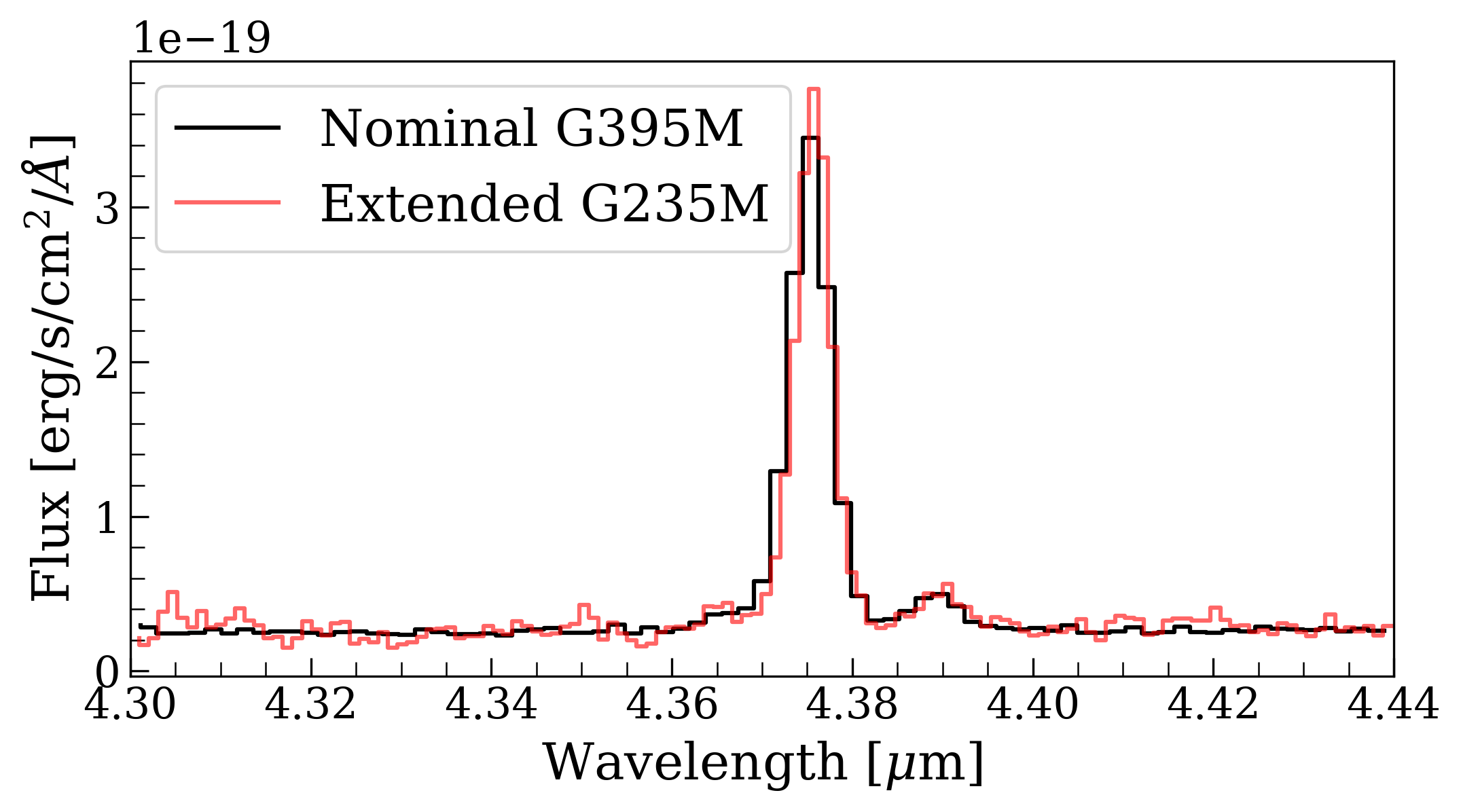}
    \caption{
    Example spectrum in the nominal range (black) and the extended spectra after calibration (red), extracted from a circular aperture of diameter 0.3$^{\prime\prime}$ centered on the brightest pixel of the star-forming galaxy DC-417567 at $z=5.67$ \citep[see][for details]{Faisst:2025}.
    The spectra show the \ha-\nii\ complex and illustrate the improvement in spectral resolution and RMS for the extended version.
    }
    \label{fig:increase_resolution}
\end{figure}

As expected from the grating equation, the spectral resolution continues to increase with wavelength, following the trend of  \citet{Jakobsen:2022}. We confirm this trend beyond the nominal wavelength range.
Figure \ref{fig:increase_resolution} shows the same \ha line from the galaxy DC-417567 at $z = 5.67$, observed in program 3045 \citep{Faisst:2025}.
We show both the extended G235M/F170LP observations and the nominal range of G395M/F290LP, highlighting the differences in resolutions and RMS.

To measure the spectral resolution beyond the nominal range, we used observations of emission lines detected both in a nominal filter (whose resolution is known and tabulated) and in the extended region and fitted them.
From the FWHM of the line measured in the nominal range ($\rm FHWM_{\rm obs, nom}$), we derived the intrinsic FWHM by correcting for instrumental broadening as $\rm FWHM_{int}$$=$$\sqrt{\rm FHWM_{\rm obs, nom}^2 - FWHM_{\rm instr, nom}^2}$. 
As a fiducial value for the resolution in the nominal range of each filter, $\rm FWHM_{\rm instr, nom}$, we used the calibrations of \citet{Shajib:2025}, who find that the in-flight spectral resolution is more than 10\% better than that estimated before launch and provided in the JWST documentation \citep{Jakobsen:2022}.
Having obtained intrinsic line width, we derived the instrumental FWHM in the extended wavelength range as $\rm FWHM_{\rm instr, ext}$$=$$\sqrt{\rm FHWM_{\rm obs, ext}^2 - FWHM_{\rm int}^2}$, where $\rm FHWM_{\rm obs, ext}$ is the measured FWHM of the line in the extended spectrum.

\begin{figure}[]
    \centering
    \includegraphics[width=0.98\linewidth]{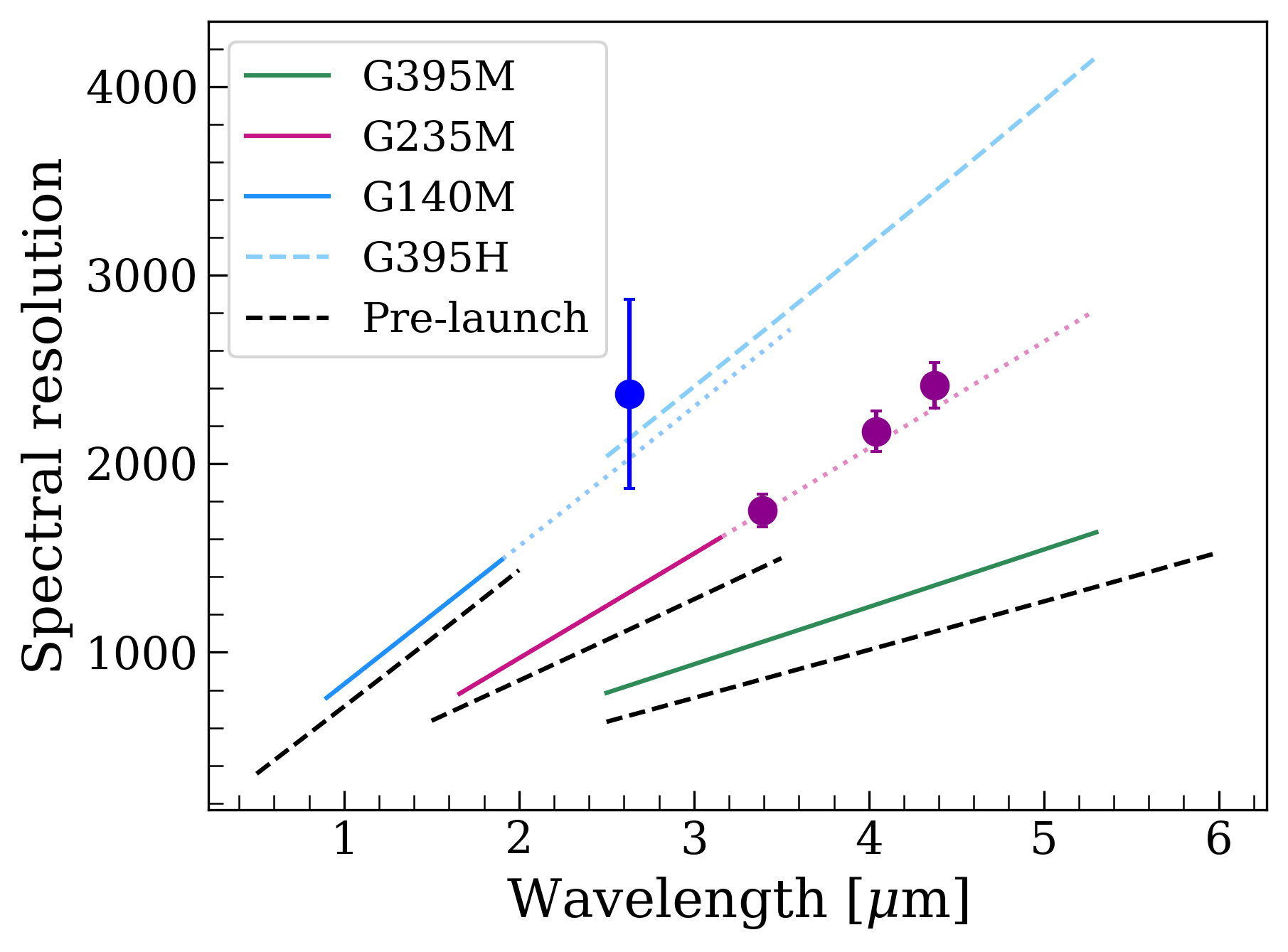}
    \caption{
    Spectral resolution as a function of wavelength.
    Dashed black lines show the JWST pre-launch expected resolution.
    Solid lines show the post-launch curves from \citealt{Shajib:2025} for G140M (blue), G235M (purple), and G395M (green).
    Dotted colored lines show the extrapolation beyond the nominal wavelength range of the post-launch curves.
    Purple and blue points show the measured resolution from the extended G235M and G140M data, respectively.
    The light-blue dashed line shows the spectral resolution curve derived for the G395H grating from \citealt{Shajib:2025}.
    }
    \label{fig:resolution}
\end{figure}

Figure \ref{fig:resolution} shows the resolution reported by the JWST documentation, that reported by \citet{Shajib:2025}, its extrapolation beyond the nominal wavelength range, and our measurement of the resolution in the extended wavelength range (colored points).
For the resolution beyond the nominal range of G235M, we used observations from the program 3045, which targeted the same galaxies with both G235M/F170LP and G395M/F290LP filter/grating combinations. We fitted the bright \oiii\ and \ha\ lines falling in the 3 -- 4.5 $\mu$m range at $z\sim 5.1-5.67$.
For G140M/F100LP, we used data from program 3435, which targets the galaxy M51, where we detect the Br$\beta$ emission line at 2.63 $\mu$m, albeit with lower S/N in the extended region. To our knowledge, no other suitable data exist in the archive with narrow emission lines observed in both G140M/F100LP and G235M/F170LP that we can exploit for our purposes.
The resolution measured in the extended region follows the same trend as the curves reported by \citealt{Shajib:2025} within the uncertainties, which we hence adopt for the remainder of this paper.
We also note that, as a result of this new reduction, observations in the extended region reach resolutions up to twice the nominal ones. Although this is advantageous, as the spectral resolution can easily reach $R$ $\sim$$2000$,  Fig. \ref{fig:resolution} shows that at the reddest wavelength our extension of G140M reaches a resolution comparable with that of the high resolution G395H data. This also implies a lower S/N and contributes to the increasing RMS at longer wavelengths.

Finally, we tested the wavelength calibration beyond the nominal range. Figure~\ref{fig:increase_resolution} shows that the centroid of the emission line falls at the same wavelengths. We checked for any discrepancies in the wavelength calibration by fitting emission lines detected in the nominal and extended gratings. Figure \ref{fig:centroids} shows the difference between the centroid positions of the same line measured in the extended and nominal ranges, which shows no systematic discrepancy between the two measurements and highlighting that the wavelength calibration in the extended range is accurate to one-quarter to one-tenth of the spectral resolution.

\subsection{Different applications }

The advantage of the data reduction presented in this work is that, in cases where two adjacent gratings, consisting of a ``blue'' filter (e.g., G140M/F100LP or G235M/F170LP) and a ``red'' filter (e.g., G235M/F170LP or G395M/F290LP), would normally be required to obtain all the spectral features of interest, this method allows the required wavelength coverage to be recovered from observations of the bluer grating alone.
This approach enables longer integrations in the bluer filter rather than splitting the exposure time between the two filters, at the cost of an increased RMS in the extended wavelength range, while having a higher spectral resolution.

Taking into account these limitations, when deep observations are performed at shorter wavelengths using a filter-grating combination, our extension naturally allows the detection of bright emission lines in the extended wavelength range.
In contrast, faint emission lines or features in the extended region are better detected using observations in the nominal filter rather than relying on the extended wavelength coverage.
In general, this approach may be useful when searching for a faint line in a blue filter, while the brighter lines in the red filters are observed simultaneously. 
This is especially relevant for dust-obscured sources, where emission lines at shorter wavelengths may be heavily attenuated.
 However, this method is also well suited for the observation of bright sources, such as QSOs or nearby galaxies, where the S/N is not a limiting factor; hence, the increase in RMS does not heavily affect the final results.

\begin{figure*}[htb!]
    \centering
\includegraphics[width=0.405\linewidth]{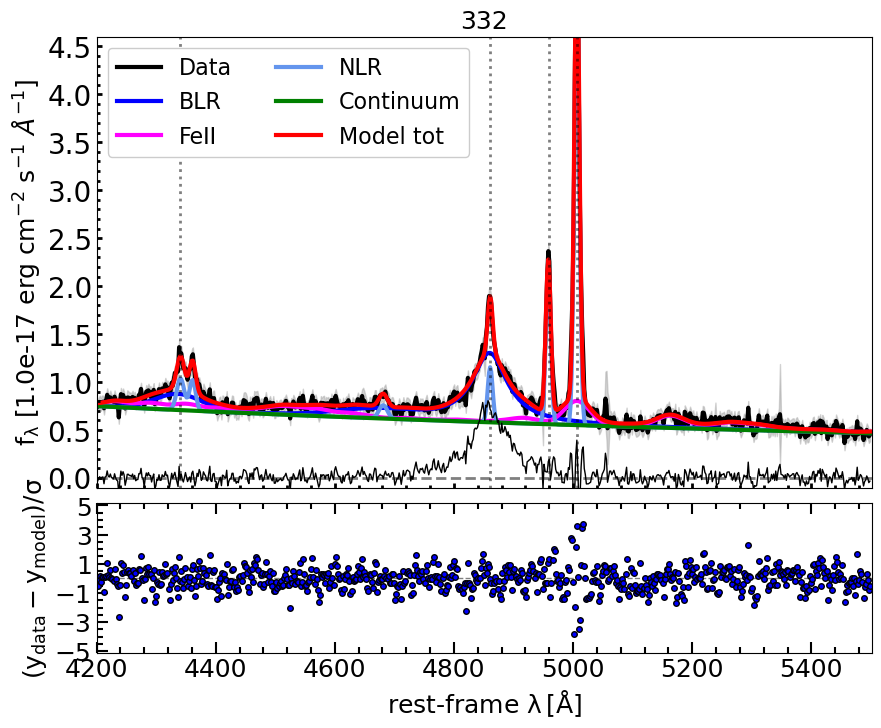}
\includegraphics[width=0.45\linewidth]{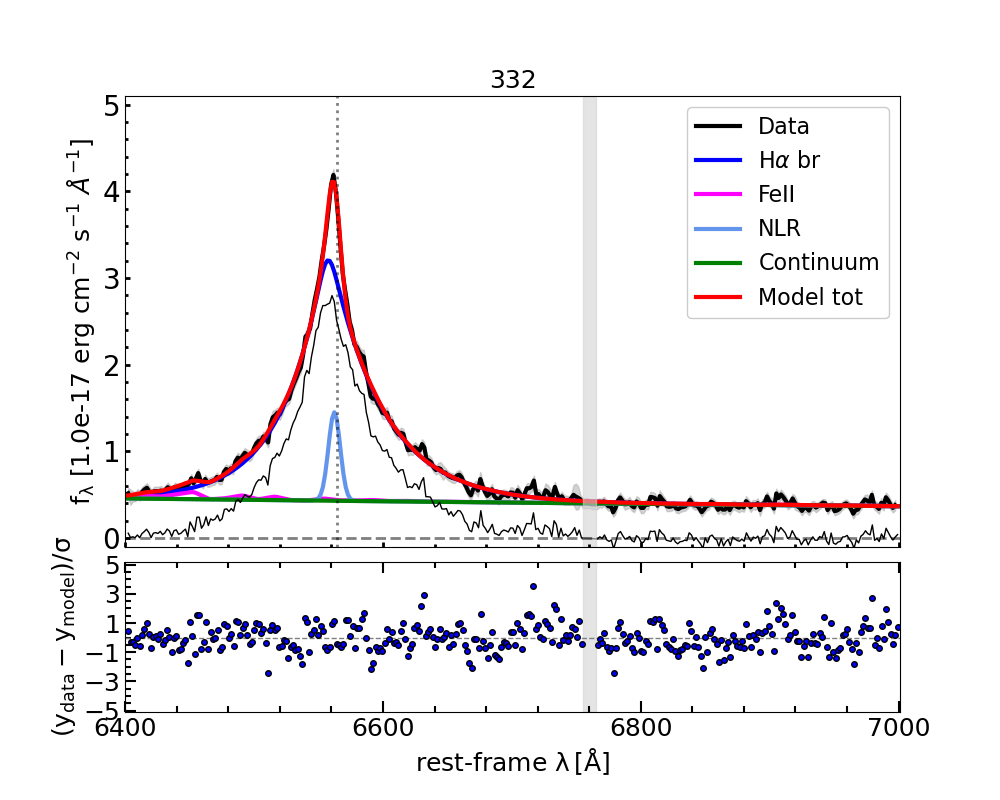}
    \caption{Rest-frame spectrum and best-fit model for RM332 for the \hb--\oiii\ line complex (left) and the \ha in the extended region (right).
    The data are shown in black and the best-fit model is shown as a solid red line.
    The total model is constructed as the sum of the BLR (blue), NLR (light blue), \Feii pseudo-continuum (magenta), and continuum emission (green). The thin black line represents the isolated broad-line profile obtained after subtraction of the best-fit model. The bottom panels show the residuals, computed as (data-model)/error.
    }
    \label{fig:spectral_fitting}
\end{figure*}

\section{Data analysis}

\label{sec:data}

In this work, we applied the procedure described above to the targets of proposal PID: 2057, which observed ten QSOs at $z\sim2$ previously targeted by RM campaigns \citep{Shen:2024}.
Among these, we focused on the five sources with available RM-based BH mass measurements \citep{Shen:2024}.
Table \ref{tab:best_fit_results} lists the targets used in this work, together with their coordinates and observation dates.
We estimated the RM-based mass from observation of the \civ (RM032, RM312, and RM539) and \mgii (RM332 and RM401) emission lines \citep{Shen:2024}.

The targets were observed with G140M/F100LP for $\sim$2400s using the 4-POINT-DITHER strategy.
We retrieved the count-rate images from the MAST archive and applied all pipeline steps described above to obtain the final datacube using a drizzle weighting method and a spaxel size of 0.05\arcsec.
We then corrected the flux by applying Eq. \ref{eq:spectrum_final}.
For all targets, the \oiii\ and \hb emission lines fall within the nominal range covered by the G140M/F100LP observations, while the \ha\ emission line can only be obtained through our extension of the G140M data.
For each cube, we removed the background by estimating it from target-free regions and extracted the spectrum from a circular aperture of 1 arcsec diameter centered on the brightest pixel to cover the extent of the point spread function \citep[$\sim 0.16''-0.18''$ at the wavelength of the Balmer emission lines, see][]{Jones:2026}, thereby maximizing the S/N and removing the effect of the ``wiggles'' visible in smaller apertures \citep{Perna:2023}. We did not perform any further aperture corrections as our aperture includes more than 95\% of the flux \citep{Jones:2026}.

To reproduce the observed spectra, we modeled them as the sum of a power-law continuum arising from the accretion disk, broad lines (FWHM $\geq$ 1000 \kms) arising from the BLR, narrow lines (FWHM $<$ 1000 \kms) arising from the narrow-line region (NLR) or the host galaxy emission, and a \Feii pseudo-continuum. 
We fitted each forbidden line with one or two components (two components when outflows were present) and each permitted line with two or three components to account for the NLR, BLR, and outflows where required.
The \hb--\oiii\ and \ha--\nii\ line complexes were modeled independently in the rest-frame wavelengths between 4200--5500\AA, and between 6200--6900\AA\ for  \hb and \ha, respectively.

We employed different spectral shapes to reproduce both the broad and narrow lines as faithfully as possible. In particular, we used both Gaussian and Lorentzian profiles for the broad lines, while Gaussian profiles were generally preferred for the narrow lines. Additionally, we constrained the kinematics of the narrow and outflow components of \hb and \oiii\ to be the same and set a 3:1 ratio between the \oiii$\lambda5007$ and \oiii$\lambda4959$ components (\citealt{osterbrock2006astrophysics}). Similarly, we fixed the amplitude of the \nii$\lambda$6583\AA to be one-third of the \nii$\lambda$6548\AA, \citealt{osterbrock2006astrophysics}. For the broad \Feii pseudo-continuum, instead of relying on an empirical template, we adopted a linear combination of theoretical templates obtained with the \textsc{cloudy} software (\citealt{ferland2017}). We then fitted each integrated spectrum using a custom Python code based on the IDL MPFIT package \citep{Markwardt:2009}, which finds the best-fit parameters by minimizing the $\chi^2$. More details on the implementation can be found in \citealt{Trefoloni:2025b}.
Since our goal was to constrain the broad-line properties, and to incorporate as much information as possible about the kinematics traced by the lines, we adopted a two-step process similar to that described in \citealt{Trefoloni:2025b}. We first performed a fit aimed at reproducing the global observed spectrum as accurately as possible. We then subtracted all best-fit components except the broad line of interest, namely \hb and \ha. We used this isolated broad-line profile to estimate the relevant quantities (e.g., flux, FWHM, and line dispersion). The result of this procedure is shown in black in Fig. \ref{fig:spectral_fitting}, where we present the best-fit model for one of the QSOs to which we applied the IFU extension procedure. The others, together with the best-fit values, are presented in Appendix \ref{appendix:spectra_qso}.

\section{Comparing different single-epoch calibrations}
\label{sec:comparing_se_masse}
After measuring the broad-line (luminosity and FWHM of \ha\ and \hb) and continuum (luminosity at 5100\AA) properties from the extended spectra (reported in Table \ref{tab:best_fit_results}), we used them to derive SE BH masses.
In this section, we compare the BH masses inferred from RM \citep{Shen:2024} with those inferred from different SE calibrations, all calibrated from local BHs using measurement of the \ha and the \hb emission lines. 
While individual GRAVITY studies of single objects have already revealed some discrepancies \citep{Abuter:2024, Gravity_z4:2025} between BH masses derived from BLR modeling and from SE scaling relations, the newly available high-redshift RM measurements provide a valuable sample to assess which SE relations yield the most reliable estimates in the early Universe.

To perform this test, we used the calibrators of \citet[][hereafter: \citetalias{Vestergaard:2006}, both the one exploiting the monochromatic luminosity at 5100\AA\ and the luminosity of the \hb emission line]{Vestergaard:2006} and \citet[][hereafter: \citetalias{Dallabonta:2020}]{Dallabonta:2020}, which are based on \hb; and \citet[][hereafter: \citetalias{Reines:2013}]{Reines:2013}, \citet[][hereafter: \citetalias{Dallabonta:2025}]{Dallabonta:2025}, and \citet[][hereafter: \citetalias{Greene:2005}]{Greene:2005}, which exploit \ha properties (FWHM and luminosity). While the typical uncertainty associated with RM \Mbh is on the order 0.3 dex (see, e.g., \citealt{Shen:2024}), that of SE \Mbh is on the order 0.2--0.5 dex (\citealt{shen2013mass, Dallabonta:2025}).

\begin{figure}
    \centering
    \includegraphics[width=0.98\linewidth]{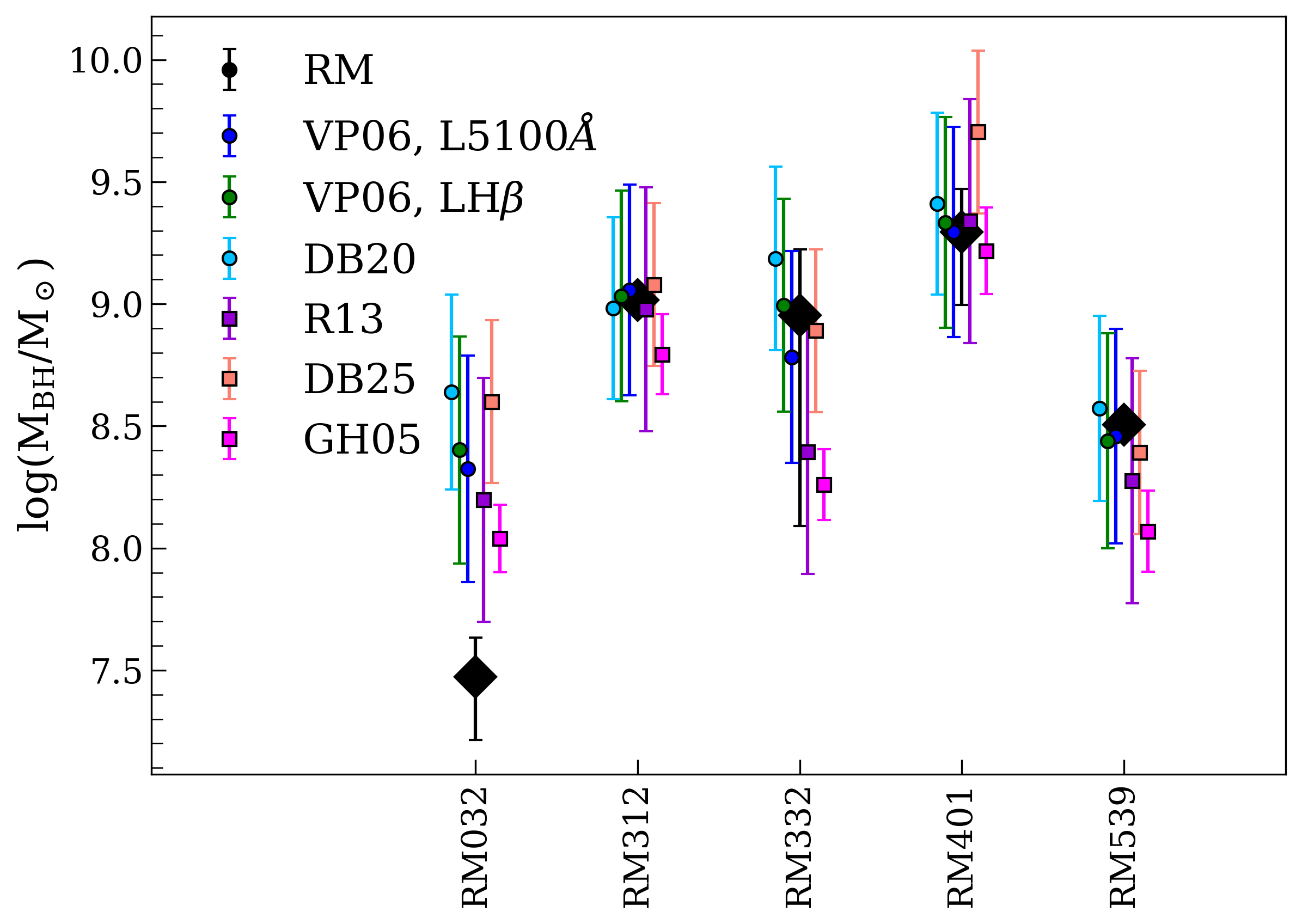}
    \caption{Comparison between the RM-estimated BH mass (large black diamond) and the SE calibrators employing the \hb emission line (VP06-L5100, VP06-L\hb, DB20; blue, green, and light-blue circles, respectively) and the SE calibrators using the \ha line (R13, DB25, GH05; purple, orange, and pink squares, respectively. }
    \label{fig:calibrators_comparison}
\end{figure}

Table \ref{tab:single_epoch_different_calibrators} lists 
the BH masses estimated from our five targets from the fitted broad emission line using the aforementioned SE calibrators, together with the literature RM-based BH mass.
Figure \ref{fig:calibrators_comparison} also shows the comparison between the different BH mass estimates.
It is evident that for some targets, such as RM312, RM401, and RM539, the majority of the calibrators agree with the RM-estimated BH mass within their uncertainties, with scatters smaller than 0.5 dex.
In the case of RM332, there is a larger scatter between the various calibrators using different emission lines, similar to what is found in other high-$z$ targets \citep{Abuter:2024, Bertemes:2025B, Marshall:2025}. 
However, RM032, the least massive BH according to the RM measurement, shows a striking difference between all calibrators and the RM-estimated mass: all SE estimates are higher than the RM value.
In particular, this analysis highlights that, using some SE calibrators, the derived BH mass can be one order of magnitude higher than that measured with the RM technique.
We note that we estimated the RM mass for RM032 using the FWHM of the \civ emission line, which can be affected by uncertainties in its modeling because it can trace non-virial motions such as outflows \citep{Denney:2012}. However, RM312 and RM539 also use the same line measure, showing good agreement between the RM and SE measures, and we find no evidence for strong outflows in the \oiii\ emission line or RM032 (see Fig. \ref{fig:spectra_1}).  

With the exception of RM032, we find that the calibrators using the \hb\ emission line show a better agreement in estimating the RM-BH mass for our sample, with the smallest deviation achieved using the  \citetalias{Vestergaard:2006} calibrations.
In contrast, the \ha-based calibrators show, in most cases, a larger scatter in the derived BH masses when different calibrators are employed; however, they are generally within 0.5 dex of the RM-based \Mbh.

To investigate the origin of the deviations in the BH mass estimates, we looked at possible correlation 
with the Eddington ratio.
This is motivated by multiple studies \citep[e.g.,][]{Du:2019,King:2024, Lupi:2024} that show that, when close to or above the Eddington limit, SE virial estimators of the BH mass might no longer be applicable.
For instance, \citet{King:2024} show that approaching the Eddington limit, the BLR could be far from virialized and instead be dominated by outflowing motions. Such behavior has also recently been observed in the BLR of a super-Eddington QSO ($\lambda_{\rm Edd}\sim 6-19$; \citealt{Gravity_z4:2025}), suggesting that standard calibrations relying on BLR virialization could fall short in this case.
\citet{Lupi:2024} point out that, in the case of super-Eddington accretion, the accretion is expected to occur via a slim disk, which would screen some of the ionizing photons and prevent them from reaching the BLR. In this case, assuming the standard $R-L$ relation would overestimate the actual BLR size, as also reported in various observations \citep{Du:2019, Abuter:2024}.
Both mechanisms would lead to an overestimation of the virial BH mass.

We estimated the Eddington luminosity as
\begin{equation}
    L_{\rm EDD} = \frac{4\pi G M_{\rm BH}m_pc}{\sigma_T}\simeq 1.26\times 10^{38} \, \bigg(\frac{M_{\rm BH}}{\rm M_{\odot}}\bigg) \, \rm erg \, s^{-1},
\end{equation}
where $M_{\rm BH}$ is the BH mass estimated from RM, G is the gravitational constant, $m_{\rm p}$ is the proton mass, $c$ is the speed of light, and $\sigma_{\rm T}$ is the Thomson scattering cross section.
We then computed the Eddington ratio ($\lambda_{\rm EDD}$) as the ratio of the observed bolometric luminosity reported in \citealt{Shen:2019}, derived from the continuum luminosity using the bolometric corrections described in \citealt{Richards:2006}, to the Eddington luminosity.
Although all other targets have Eddington ratios between 0.03 and 0.12, RM032 has $\lambda_{\rm EDD}=0.8\pm0.3$.

In the case where the RM measurements of RM032 are not affected by non-virial motion traced by the \civ line, the results presented in this work suggest that, at high redshift, most SE calibrators overestimate the masses of low-mass, rapidly accreting BHs. 
If confirmed, this could reconcile recent results from JWST-discovered BLAGN, in which the BH masses appear comparable to the stellar masses of their hosts \citep{Kokorev:2023,Juodzbalis:2024}.
However, our current sample includes only one target with log(M$_{\rm BH}$/M$_\odot$) $< 8$, so larger statistics are needed to confirm this trend.

\begin{table*}
    \centering
\caption{Comparison of different BH mass estimates for the five targets in this work.
}
\label{tab:single_epoch_different_calibrators}
    \begin{tabular}{|c|c|c|c|c|c|c|}
    \hline\hline
         & Calibrator & RM032&  RM312&  RM332&  RM401& RM539\\ \hline
         \textit{z}& -  &  1.714 &  1.92 &  2.58 &  1.822 & 2.27 \\ \hline
         $\log  (M_{\rm BH, RM} / M_\odot)   $& - &  7.47 $^{+0.22}_{-0.26}$&  9.01$^{+0.03}_{-0.05}$&  8.95$^{+0.27}_{-0.26}$&  9.29$^{+0.23}_{-0.30}$& 8.50$^{+0.05}_{-0.02}$\\ \hline
         $\log  (M_{\rm BH, SE} / M_\odot)   $& VP06, FWHM(\hb), L5100 & 8.33 $\pm$ 0.46&  9.06 $\pm$ 0.43&  8.78 $\pm$ 0.43&  9.29 $\pm$ 0.46& 8.45 $\pm$ 0.44\\ \hline
         $\log  (M_{\rm BH, SE} / M_\odot)   $& VP06, FWHM(\hb), L\hb & 8.40 $\pm$ 0.47&  9.03 $\pm$ 0.43&  8.99 $\pm$ 0.43&  9.33 $\pm$ 0.43& 8.44 $\pm$ 0.44\\ \hline
         $\log  (M_{\rm BH, SE} / M_\odot)   $& DB20, FWHM(\hb), L\hb & 8.64 $\pm$ 0.39&  8.98 $\pm$ 0.37&  9.18 $\pm$ 0.37& 9.40 $\pm$ 0.37&8.57 $\pm$ 0.38\\ \hline

 $\log  (M_{\rm BH, SE} / M_\odot)   $& R13, FWHM(\ha), L\ha& 8.19 $\pm$ 0.5& 8.97 $\pm$ 0.5& 8.39 $\pm$  0.5& 9.34 $\pm$ 0.5&8.28 $\pm$  0.5\\\hline
 $\log  (M_{\rm BH, SE} / M_\odot)   $& DB25, FWHM(\ha), L\ha& 8.60 $\pm$  0.33& 9.03$\pm$ 0.33& 8.89 $\pm$ 0.33& 9.70 $\pm$ 0.3&8.39 $\pm$ 
 0.3\\\hline
  $\log  (M_{\rm BH, SE} / M_\odot)   $& GH05, FWHM(\ha), L\ha& 8.04 $\pm$ 0.14& 8.79  $\pm$ 0.17& 8.26  $\pm$ 0.14& 9.2  $\pm$ 0.17&8.06 $\pm$ 0.15\\\hline

    \end{tabular}
\tablefoot{The first and second columns list the measurement and the calibrators used to derive it, respectively. Columns 2-7 list all BH mass estimated for each target. }    
    
\end{table*}

\section{Calibrating the single-epoch relation at z$\sim$2}

\label{sec:calibrating_se}
In the previous section, we tested various SE calibrators to estimate BH masses for high-$z$ BHs.
Since all the SE calibrators have so far been derived based on local AGN, we set out to derive a first SE calibration at high-z using the five targets above. We additionally included targets with dynamical BH estimates from GRAVITY at $z\sim2$ and $z\sim4$ from \citealt{Abuter:2024} and \citealt{Gravity_z4:2025}, another target with a BH mass inferred from RM and \hb detection at $z\sim4$ \citep{Saturni:2018}, and the $z\sim 7$ LRD with a kinematically measured BH mass \citep{Juodzbalis_2025_dynamical} and broad \ha and \hb analysis \citep[][respectively]{Deugenio:2025_QSO1, Ji:2025}.
In particular, we fitted the relation

\begin{equation}
\label{eq:calibration}
    \log({\rm M_{BH}/M_\odot)} = a + b  [ \log(L) - 42] + c  [\log({\rm FWHM}) - 3],
\end{equation}
where $\rm M_{BH}$ is the mass from RM campaigns (or from BLR modeling for the GRAVITY targets), L is the luminosity (of the \ha, \hb emission line or monochromatic continuum emission at 5100\AA in units of erg s$^{-1}$), and FWHM is the FWHM of the \ha or \hb broad emission lines in units of \kms obtained from our fitting. 
We simultaneously fitted the parameters $a$, $b$, and $c$, together with the intrinsic scatter $\sigma_{\rm int}$. 
We performed the fit using a Markov chain Monte Carlo approach, which allowed us to explore the posterior distributions of the parameters and account for the measurement uncertainties. We adopted uniform priors on $a$, $b$, $c$, and $\sigma_{\rm int}$, ran 50 chains with 100,000 steps to ensure convergence, and discarded the initial 10,000 as the initial burn-in phase. 
Figures \ref{fig:corner_calibrations_ha}, \ref{fig:corner_calibrations_hb}, and  \ref{fig:corner_calibrations_l5100} show the posterior distributions of the parameters, while Table \ref{tab:values_se_calibrators} reports the best-fit values for each calibrator and their uncertainties estimated as the 16\textsuperscript{th} and 84\textsuperscript{th} percentiles of the posterior distribution.

\begin{figure}[ht!]
    \centering
    \includegraphics[width=0.95\linewidth]{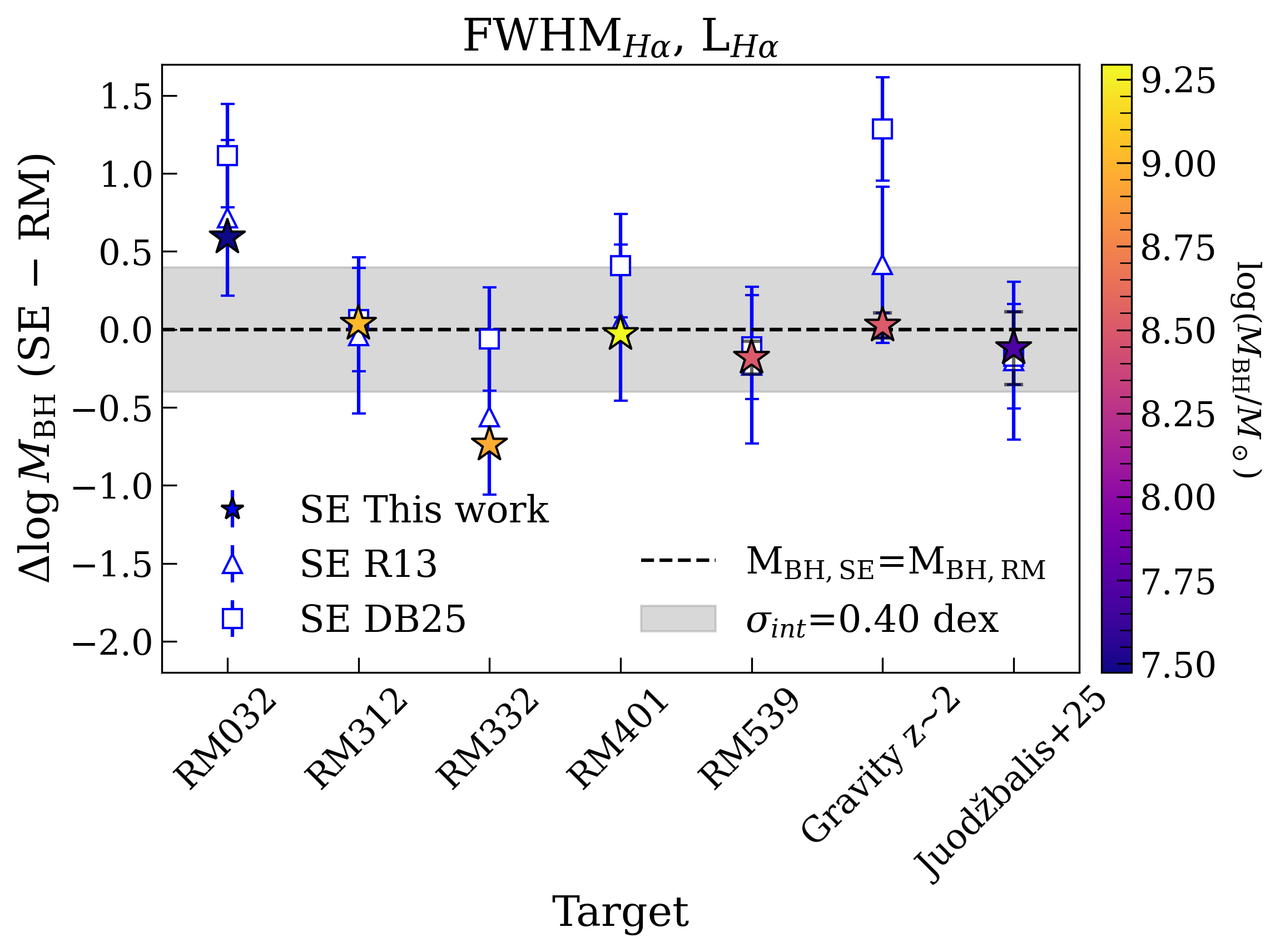}
    
    \includegraphics[width=0.95\linewidth]{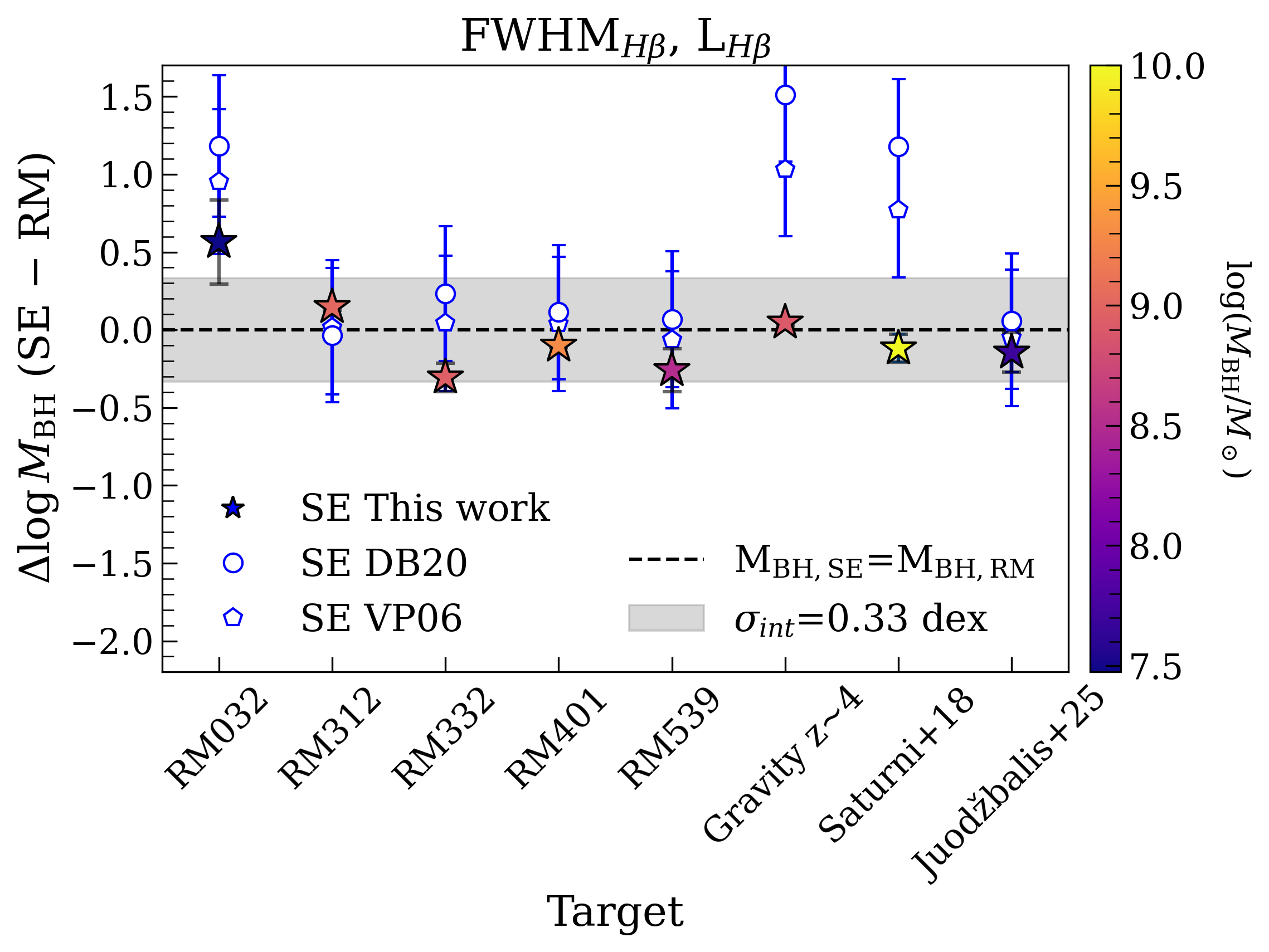}
    \includegraphics[width=0.95\linewidth]{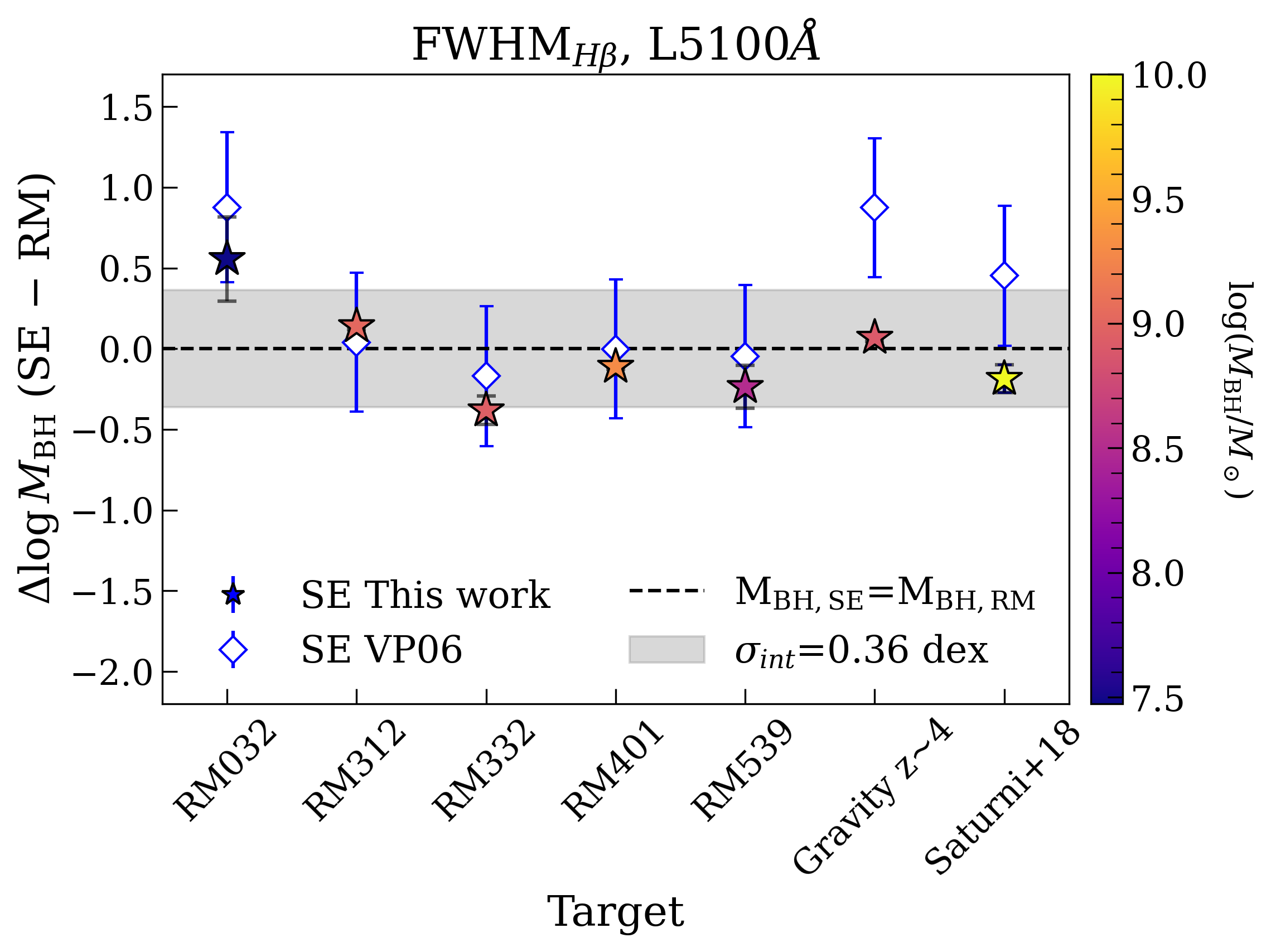}
    \caption{
    Comparison between RM (or kinematic) BH masses and SE BH masses derived using the relation from this work (stars) and other commonly used SE relations. From top to bottom: Differences in SE $\mathrm{M_{BH}}$ estimated from $L_{\rm H\alpha}$ and FWHM$_{\rm H\alpha}$, $L_{\rm H\beta}$ and FWHM$_{\rm H\beta}$, and $L_{5100\AA}$ and FWHM$_{\rm H\beta}$, respectively. Each target is color-coded by its RM $\mathrm{M_{BH}}$. Error bars on our SE estimates show only statistical uncertainties from measured FWHM and luminosities, excluding intrinsic scatter and calibration errors $a$, $b$, and  $c$. The shaded gray area shows the intrinsic scatter of our relation.}
    \label{fig:calibrations}
\end{figure}

Figure \ref{fig:calibrations} shows the comparison between the BH mass from the RM mapping and kinematics measure and the SE mass estimated from our calibration for the \ha and \hb emission line measurements, along with the other typical calibrators used for SE M$_{\rm BH}$ estimation.
Our calibrations give \Mbh values consistent with the other calibrators for our sources, within the large uncertainties due to the small number of points fitted and the small range of BH masses in our sample, spanning only 2--2.5 dex.
We recover the mass of RM032 within 1$\sigma$ and also reproduce the mass from the GRAVITY targets, both at $z\sim2$ using \ha\ and especially at $z\sim4$ using \hb, for which all the other SE calibrators yield BH masses an order of magnitude higher \citep{Gravity_z4:2025} than those expected from the BLR kinematic estimate.

\begin{figure}[ht!]
    \centering
    \includegraphics[width=0.98\linewidth]{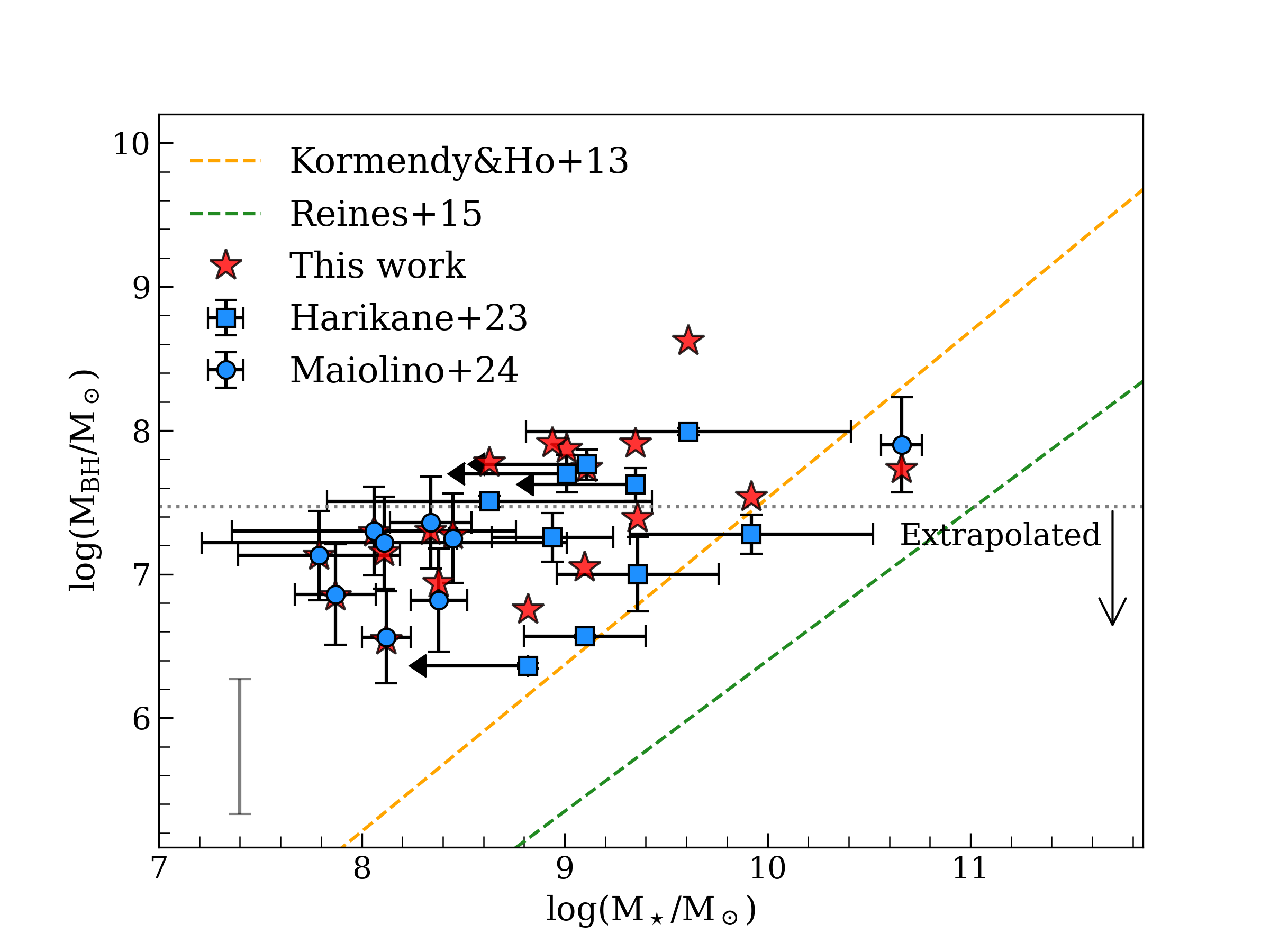}
    \caption{
    \Mbh--stellar mass relation for the high-z BHs studied by \citet{Harikane:2023} and \citet{Maiolino:2024}.
    Blue points show the \Mbh values derived by \citet{Harikane:2023} and \citet{Maiolino:2024} using SE relations from \citetalias{Greene:2005} and \citetalias{Reines:2013}, respectively.
    Red stars show \Mbh values obtained using the SE H$\alpha$-based equation derived in this work. The typical systematic uncertainties on SE-\Mbh are shown as a gray line in the lower-left corner.
    The dashed yellow and green lines show the local \Mbh-M$_{\star}$ relation from \citealt{Kormendy:2013} and \citealt{Reines:2015}, respectively.
    The dashed gray line marks the regime below which our calibration is extrapolated.
}
    \label{fig:lrd_calibration}
\end{figure}

We also applied our calibration to the newly discovered JWST high-$z$ AGN population, using the H$\alpha$ fluxes and FWHM values reported in \citet{Harikane:2023} and \citet{Maiolino:2024}. We note that these sources span redshifts between four and seven, and we have only one target at such a high redshift, while all the others have $z<4$.
The revised BH-mass estimates are consistent within the uncertainties with the values reported in the literature, which were derived using the calibrations of \citet{Greene:2005} and \citet{Reines:2013}. Even with our new calibration, the JWST high-$z$ AGN population lies above the local relation (Fig.~\ref{fig:lrd_calibration}), indicating that the detection of ``overmassive'' BHs does not depend on the specific BH-mass estimator adopted. However, we emphasize that our relation is calibrated only for $M_{\rm BH} \gtrsim 10^{7.5}$M$_\odot$, and extrapolating it to lower masses may be unreliable and could lead to bias.

Finally, we tested our relations on a larger sample of local QSOs from \citet{wu2022catalog}. In Appendix \ref{appendix:SE_relations_test} we show the differences between the BH mass inferred from our relations and those inferred using the SE relations of \citetalias{Vestergaard:2006} and \citetalias{Reines:2013}.
We show that, using our relation based on \hb, \Mbh\ is typically overestimated or consistent with standard calibrators at the low-luminosity end, while it is on average underestimated by up to 0.4 dex at the highest luminosities when compared with the \citetalias{Vestergaard:2006} calibration.
This trend is significantly less pronounced when using \ha, although it does not disappear entirely.
Larger, unbiased samples spanning a broad range of luminosities and BH masses are required to draw more robust conclusions and address the reasons for these discrepancies.

\begin{table}
    \caption{Best-fit parameters of the three relations using the \ha\ line, the \hb FWHM and luminosity, and the \hb\ FWHM and the continuum luminosity at 5100\AA.}

    \centering

    \begin{tabular}{|c|c|c|c|c|}\hline

         Calibrators&  $a$&  $b$& $c$ & $\sigma_{int}$\\\hline
         L$_{\rm H\alpha}$, FWHM$_{\rm H\alpha}$& $6.70^{+0.52}_{-0.57}$ & $0.28^{+0.22}_{-0.16}$  & $2.38^{+0.79}_{-0.81}$  & 0.40\\\hline
         L$_{\rm H\beta}$, FWHM$_{\rm H\beta}$& $6.17^{+0.56}_{-0.55}$ & $0.28^{+0.13}_{-0.12}$ &  $3.11^{+0.82}_{-0.85}$ &  0.33\\\hline
 L$_{5100 \AA}$, FWHM$_{H\beta}$& $5.78^{+0.71}_{-0.51}$ & $0.25^{+0.12}_{-0.12}$  &  $3.00^{+0.79}_{-0.95}$ & 0.35\\ \hline
    \end{tabular}
    \label{tab:values_se_calibrators}
\end{table}

\section{Summary and conclusions}

\label{sec:conclusions}

In this work, we present and release a modification of the JWST science calibration pipeline for NIRSpec IFU observations that more than doubles the wavelength coverage achievable with the medium-resolution (R1000) G140M/F100LP and G235M/F170LP configurations (see Table~\ref{tab:new_coverage}). We also provide the corresponding reference and calibration products as machine-readable files, enabling the community to run the data-reduction workflow directly and obtain final flux-calibrated datacubes over the extended wavelength range.
The extended coverage maximizes the scientific return from archival datasets and offers new opportunities for upcoming JWST observing programs.
We note that spectra in the extended wavelength range exhibit a modest increase in the RMS level (up to a factor of two to three compared with observations obtained in the nominal filter). As a result, the method is best suited for bright targets or for detecting longer-wavelength spectral features (e.g., rest-frame optical and near-IR lines) as a secondary objective in observing programs with long exposure times that primarily focus on detecting fainter, shorter-wavelength emission, such as the rest-frame UV lines. The extended spectra reach a resolution of up to R~$\sim2500$, approximately twice that of the spectra in the nominal range and comparable to that of the high-resolution gratings (R~$\sim$~2700).

We used our new data reduction to recover the \ha emission line of a sample of five QSOs at $z\sim 2$, for which only \hb\ and \oiii\ fall within the nominal wavelength range. 
We selected these targets because they have independent BH mass measurements from RM campaigns \citep{Shen:2024} based on the \mgii or \civ emission lines.
Our extension of the G140M/F100LP data also allowed us to obtain the \ha\ flux and FWHM, allowing a direct comparison between RM BH masses and the most commonly used SE calibrators at high redshift, which typically employ the \ha and \hb emission lines. All SE relations that we used were calibrated locally, but are routinely used to infer BH masses up to $z\sim9$ \citep[e.g.,][]{Tripodi:2024, Taylor:2025}.

We find that the majority of the calibrators provide a BH mass consistent within the uncertainties with the RM-based values at the high-mass end. 
However, none of the calibrations reproduce the QSO with the lowest RM-based mass ($\log M_{\rm BH}/M_\odot \sim 7.5$). 
In this case, several SE estimators predict BH masses higher by more than an order of magnitude. Notably, this BH also has the highest Eddington ratio of our sample, reaching close to the super-Eddington limit ($\lambda_{\rm EDD} = 0.8\pm0.3$).
Although based on our limited sample, these results underline the possibility of a bias in BH mass estimates for low-mass, highly accreting BHs, highlighting the need for kinematic and RM measurements at lower masses and higher redshifts to validate the SE scaling relations across the range of BH masses and cosmic times.

Finally, we used our new measurements to provide the first SE mass estimators based on \ha\ and \hb\ calibrated at high redshift, by combining them with dynamical BH masses obtained at high redshift from GRAVITY.
Although our limited sample of a few targets with $\log(M_{\rm BH}/M_\odot) \sim 7.5$–$10$ shows no evidence for evolution in the SE BH mass scaling relation at $z \sim 2$ within the large uncertainties, a larger sample spanning a broader mass range is needed to draw a more statistically robust conclusion.

\section*{Data availability}
This work is based on observations made with the NASA/ESA/CSA James Webb Space Telescope. Some of the data were obtained from the Mikulski Archive for Space Telescopes at the Space Telescope Science Institute, which is operated by the Association of Universities for Research in Astronomy, Inc., under NASA contract NAS 5-03127 for JWST. These observations are associated with programs 1536, 1537, 1538, 2057, 2186, 2654, 3045, 3435, and 6645.
The modified reference files, along with documentation detailing the required modifications to the official pipeline and the Python scripts used for the flux calibration, are publicly available on GitHub (\url{https://github.com/eleonoraparlanti/nirspecIFU-extended}), allowing users to reproduce and apply the extended pipeline version described in this work.

\begin{acknowledgements}
BT, SC, GV, SZ acknowledge support by European Union’s HE ERC Starting Grant No. 101040227 - WINGS.
MP acknowledges support through the grants PID2021-127718NB-I00, PID2024-159902NA-I00, and RYC2023-044853-I, funded by the Spain Ministry of Science and Innovation/State Agency of Research MCIN/AEI/10.13039/501100011033 and El Fondo Social Europeo Plus FSE+.
GT, acknowledges funding by the European Union (ERC Advanced Grant GALPHYS, 101055023).
H\"U acknowledges funding by the European Union (ERC APEX, 101164796). Views and opinions expressed are however those of the authors only and do not necessarily reflect those of the European Union or the European Research Council Executive Agency. Neither the European Union nor the granting authority can be held responsible for them.
\end{acknowledgements}

\bibliographystyle{aa}
\bibliography{aa}

\appendix

\section{Extended S-flat}
\label{appendix:extended_sflat}

\begin{figure*}[h!]
    \centering
    \includegraphics[width=0.98\linewidth]{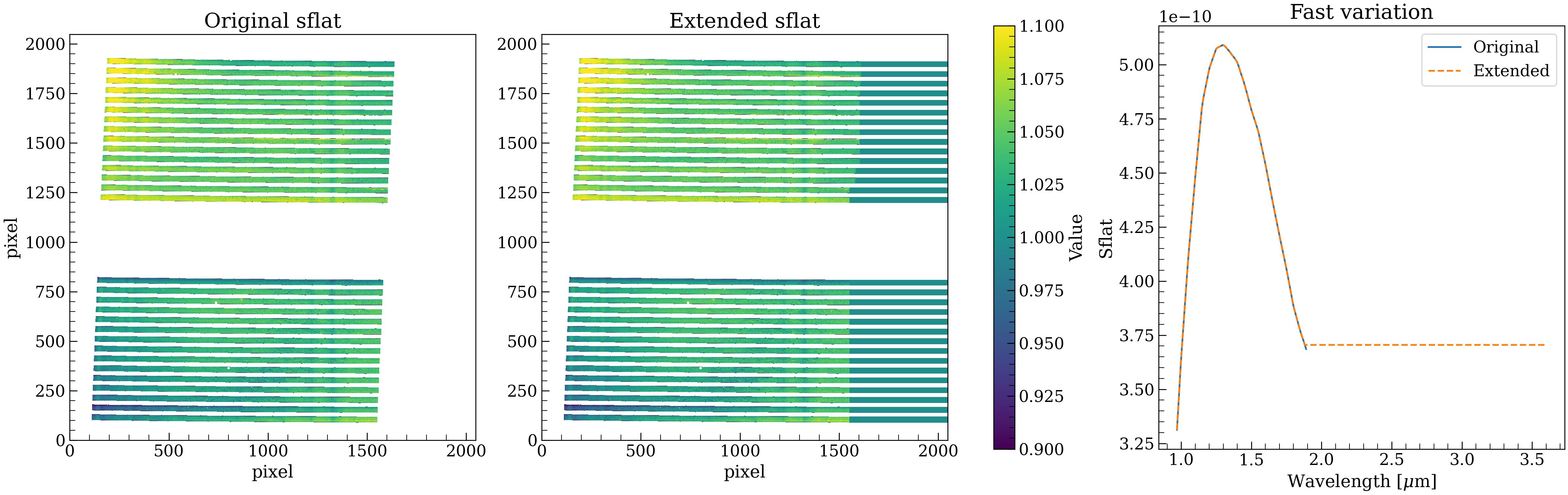}
    \caption{Detector response of S-flat for NRS1 before (left) and after (center) the extension.
            In the right panel, we show the original "FAST\_VARIATION" table as a function of wavelength in blue and the extended one in orange. }
    \label{fig:sflat_image}
\end{figure*}

In Fig. \ref{fig:sflat_image} we show the comparison between a nominal and an extended S-flat file.

\section{Flux calibration G235M/F170LP}
\label{appendix:fluxcalibration235m}

\begin{figure}[htb!]
    \centering
    \includegraphics[width=0.98\linewidth]{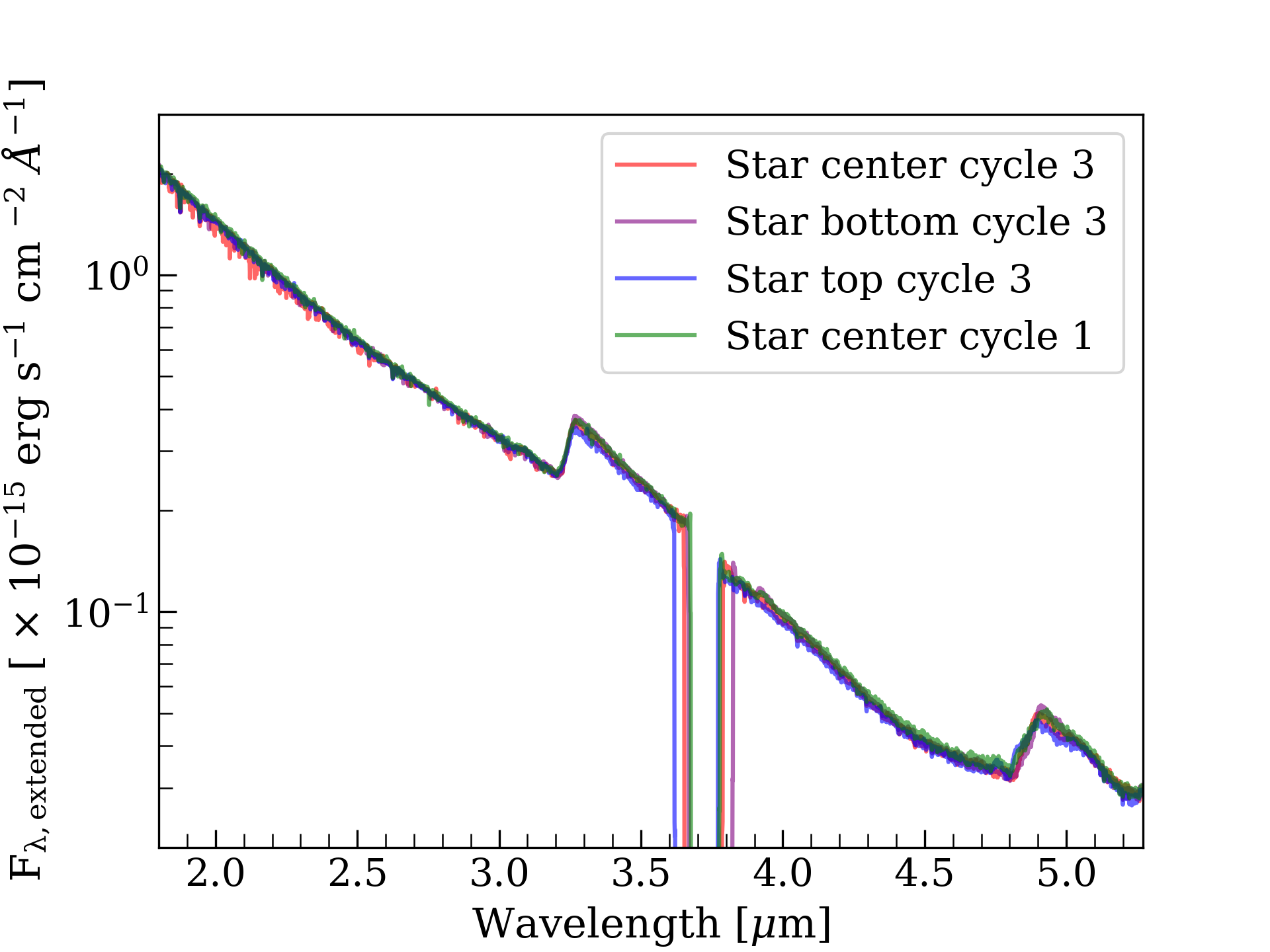}
    \caption{Observed flux in the nominal wavelength range of the extended G235M/F170LP filter for the star P330-E observed as part of the proposals 1538 and 6645.
    The curves of different colors show the same flux for the star observed in cycle 1 (green), in cycle 3 at the center of the FOV (red), and in the bottom and top parts of the FOV (blue and purple, respectively).
    }
    \label{fig:std_star_different_epoch_g235m}
\end{figure}

Following the same approach as for the G140M/F100LP, we extended the G235M/F170LP up to 5.3 $\mu$m, that is, the wavelength range encompassed by the G235M/F170LP and the G395M/F290LP.
Since the flats are extended with ones the reduced spectrum is contaminated by the presence of the second and third order spectra, we require observations of the same targets with the G235M/F170LP and the G395M/F290LP to compute the correction factors $\tilde{\alpha}(\lambda)$, $\tilde{\beta}(\lambda)$ and $k(\lambda)$ to calibrate the flux.
Before computing them, we verified any possible dependency on the position of the source on the detector and the observation time.
Figure \ref{fig:std_star_different_epoch_g235m} shows the same as Fig. \ref{fig:std_star_different_epoch}, but for the extended G235M. Even in this case, by exploiting various observations of the same standard star, we noticed no major dependence on the cycle in which it was observed, nor on the position where the spectrum fell on the detector.
In Table \ref{tab:claibration_target}, we report the relevant information, such as program ID, target name, grating/filter combination, exposure time, and observation date for each dataset used for the flux calibration of the extended region. We also report the relevant information for the target used to verify the wavelength calibration and to measure the spectral resolution in the extended region.

\begin{table}[ht!]
\footnotesize
\caption{Datasets used for calibration.}

    \centering
    \begin{tabular}{|c|c|c|c|c|}
    \hline
    PID&    Target &Grating/Filter& \makecell{Exp. \\ time(s)} &Obs. date \\
        \hline
 \multicolumn{5}{|c|}{Flux calibration}\\
    \hline
     1536& J1743045&G140M/F100LP&525&Sep 3, 2023\\
     1536& J1743045&G235M/F170LP&641&Sep 3, 2023\\
     1536& J1743045&G395M/F290LP&2100&Sep 3, 2023\\
     1537&  G191-B2B&G140M/F100LP&933&Sep 15, 2023\\
     1537&  G191-B2B&G235M/F170LP&1342&Sep 15, 2023\\
     1537&  G191-B2B&G395M/F290LP&5135&Sep 15, 2023\\
     1538& P330E&G140M/F100LP&233&Aug 7, 2022\\
     1538& P330E&G235M/F170LP&291&Aug 7, 2022\\
     1538& P330E&G395M/F290LP&700&Aug 7, 2022\\
     6645& P330E&G140M/F100LP&291&Jun 4/6, 2025\\
     6645& P330E&G235M/F170LP&408&Jun 4/6, 2025\\
     6645& P330E&G395M/F290LP&1108&Jun 4/6, 2025\\
     2654& SDSSJ0749&G140M/F100LP&3413&Nov 23, 2022\\
     2654& SDSSJ0749&G235M/F170LP&3413&Nov 23, 2022\\
     2654& SDSSJ0841&G140M/F100LP&3413&Apr 20, 2023\\
     2654& SDSSJ0841&G235M/F170LP&3413&Apr 20, 2023\\
     
     2186& MRK-273&G235M/F170LP&257&Mar 1, 2023 \\
     2186& MRK-273&G395M/F290LP&300&Mar 1, 2023 \\

     2186& UGC-5101&G235M/F170LP&300&Feb 15, 2023  \\
     2186& UGC-5101&G395M/F290LP&600&Feb 15, 2023  \\
     
     2186& IRAS-10565&G235M/F170LP&257& Nov 26, 2022 \\
     2186& IRAS-10565&G395M/F290LP&300& Nov 26, 2022 \\
     \hline
 \multicolumn{5}{|c|}{Wavelength calibration and resolution}\\
      \hline

     3435& M51-NE &G140M/F100LP&816&    Apr 2, 2024 \\
     3435& M51-NE &G235M/F170LP&816&    Apr 2, 2024\\

     3045& DC-417567 &G235M/F170LP&7236&        May 26, 2024 \\
     3045& DC-417567 &G395M/F290LP&7352&        May 26, 2024\\
     3045& DC-873321 &G235M/F170LP&1429&        Apr 26, 2024 \\
     3045& DC-873321 &G395M/F290LP&7119&        Apr 26, 2024\\
    \hline

\end{tabular}
\tablefoot{In the top part of the table the target used for flux calibration, in the bottom part, the ones used for the wavelength calibration.
The columns represent the program ID, Target Name, the grating-filter configuration used, the exposure time, and the observation date.}
    \label{tab:claibration_target}
\end{table}

\begin{figure*}[htb]
    \centering
    \includegraphics[width=0.9\linewidth]{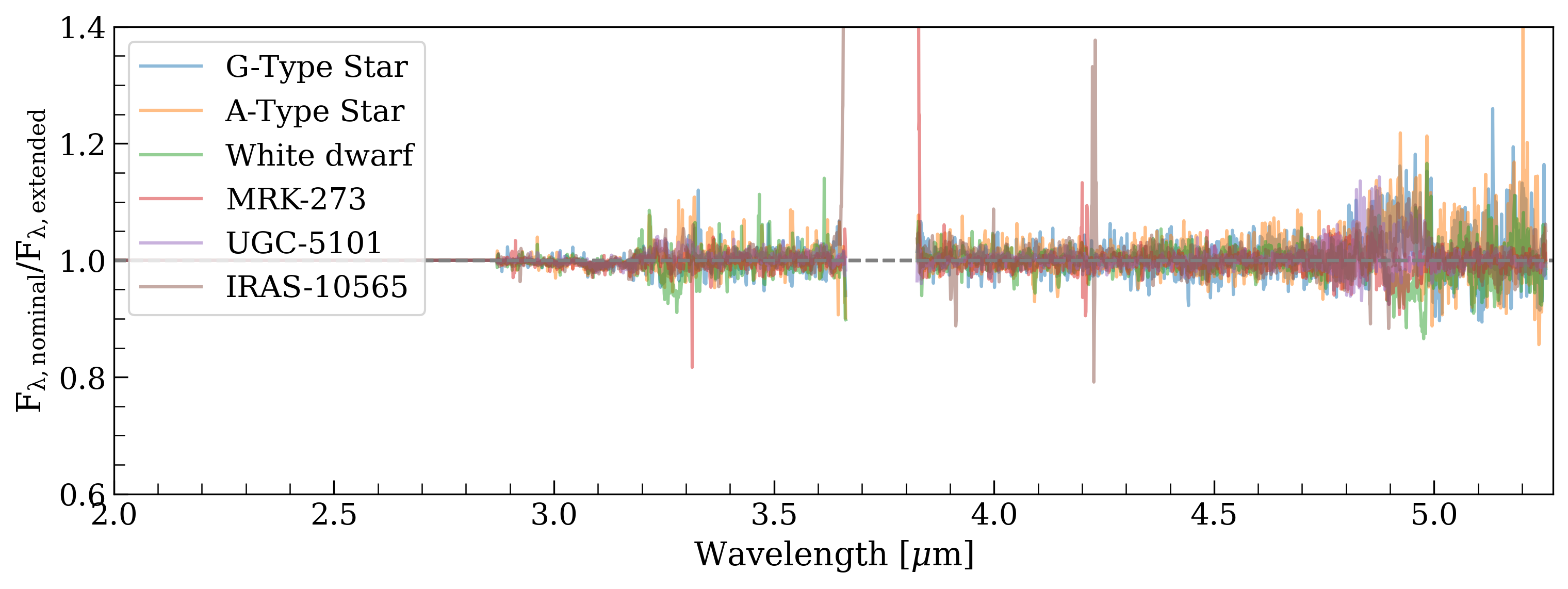}
    \caption{Ratio between the observed flux in the nominal wavelength range of the  G235M/F170LP and G395M/F290LP and the flux observed with the extended G235M/F170LP.
    }
    \label{fig:second_order_correction_g235m}
\end{figure*}

Figure \ref{fig:second_order_correction_g235m} shows the same as Fig. \ref{fig:second_order_correction}, but for the extended G235M, highlighting that flux calibration remains below 20\%.

\newpage
\section{Wavelength calibration}
\label{appendix:wavelength_calibration}

\begin{figure}[ht!]
    \centering
    \includegraphics[width=0.98\linewidth]{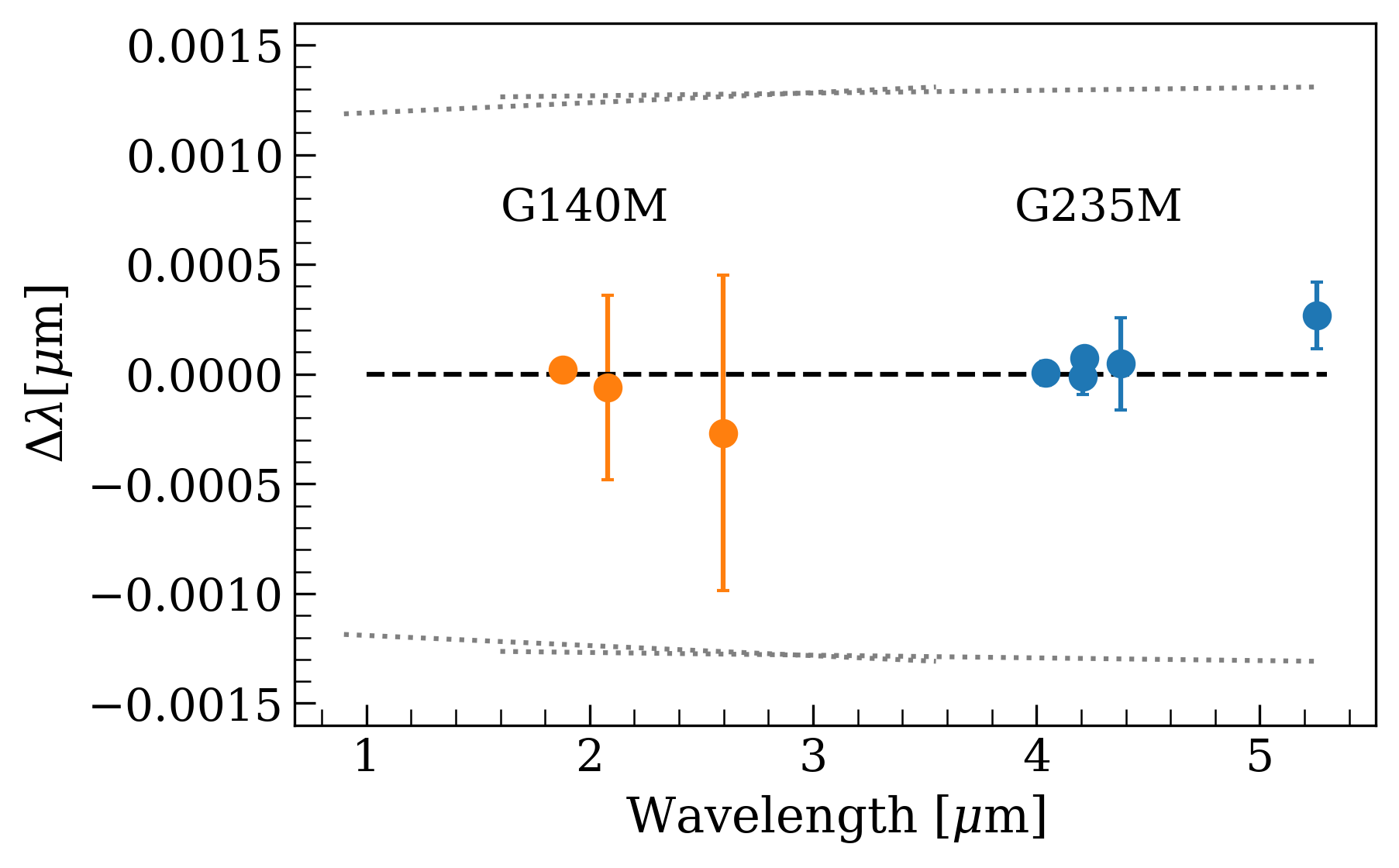}
    \caption{
    Comparison of the line centroids measured
in the extended data and grating in the nominal range.
In orange and blue, we report the measurements for the extended G140M/F100LP and G235M/F170LP, respectively.
The dotted lines represent the spectral resolution from \citet{Shajib:2025}.
    }
    \label{fig:centroids}
\end{figure}

We also verify the wavelength calibration beyond the nominal wavelength range by measuring the centroid of various emission lines that are detected both in the extended spectra and in the nominal wavelength range. Figure \ref{fig:centroids} reports the difference between the two measured centroids, highlighting that the observed variations are well within the spectral resolution typically compatible with zero.

\section{Spectral fitting best-fit results}
\label{appendix:spectra_qso}
\begin{figure*}[ht!]
    \centering
    \includegraphics[trim = 0 0 0 0cm, clip, width=0.38\linewidth]{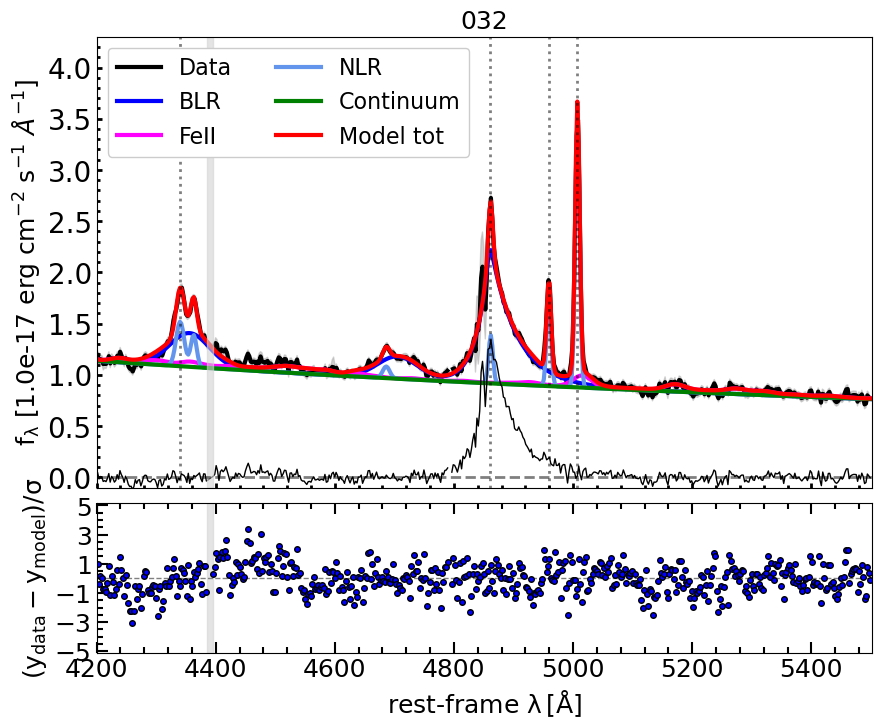}
    \includegraphics[trim = 0 0 0 1.6cm, clip, width=0.43\linewidth]{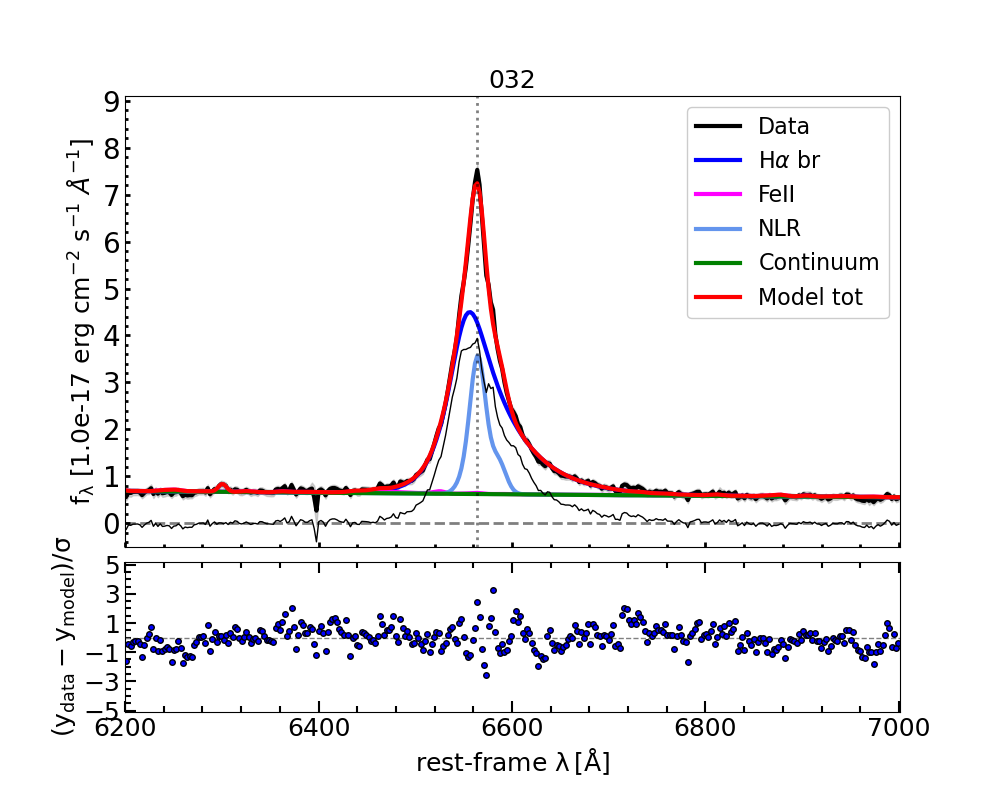}
\includegraphics[trim = 0 0 0 0cm, clip, width=0.38\linewidth]{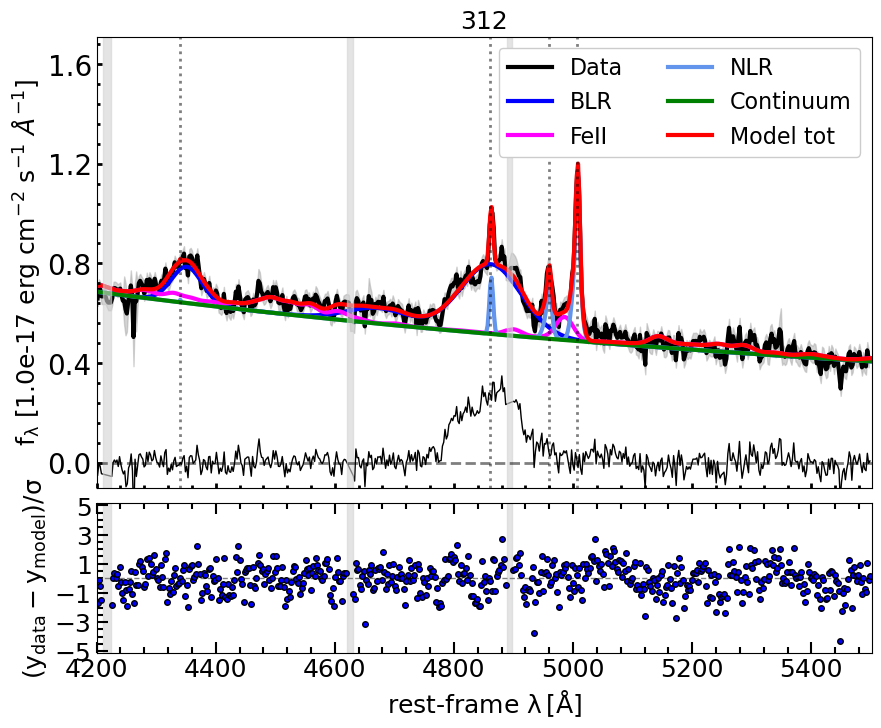}
    \includegraphics[trim = 0 0 0 1.4cm, clip, width=0.43\linewidth]{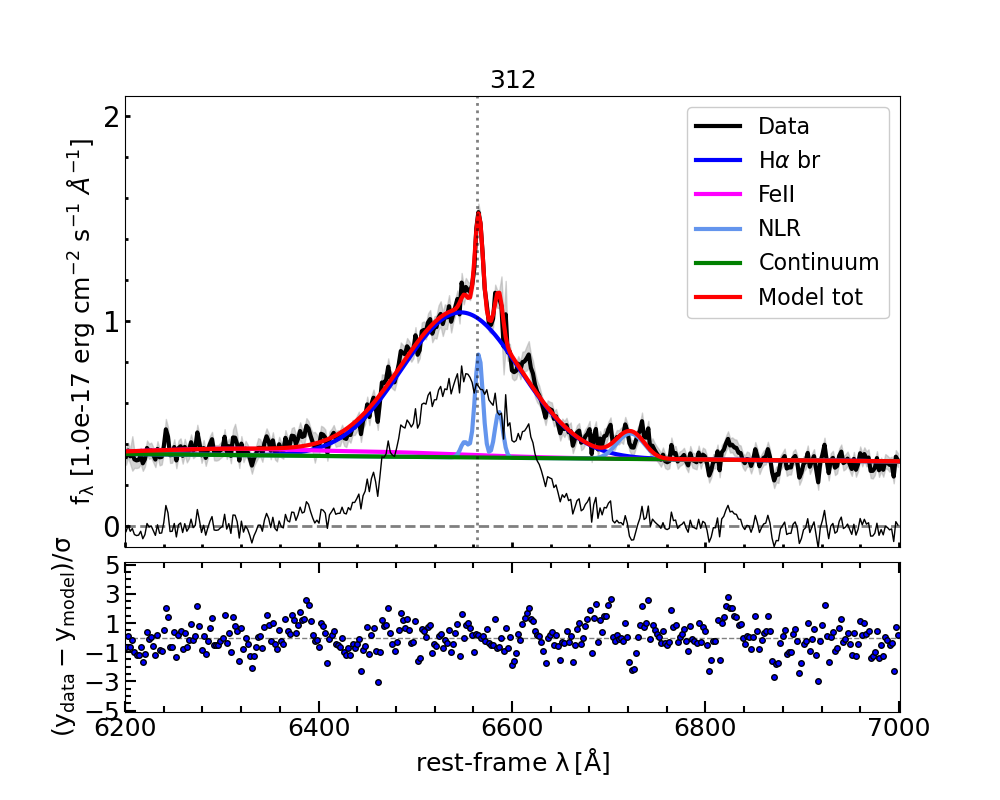}
    \includegraphics[trim = 0 0 0 0cm, clip, width=0.38\linewidth]
{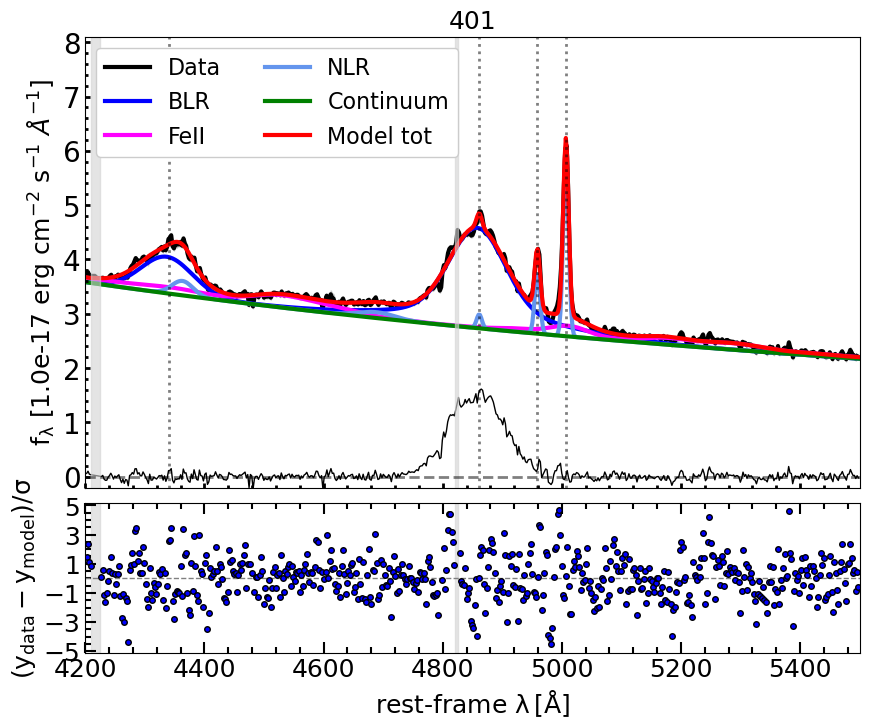}
    \includegraphics[trim = 0 0 0 1.4cm, clip, width=0.43\linewidth]{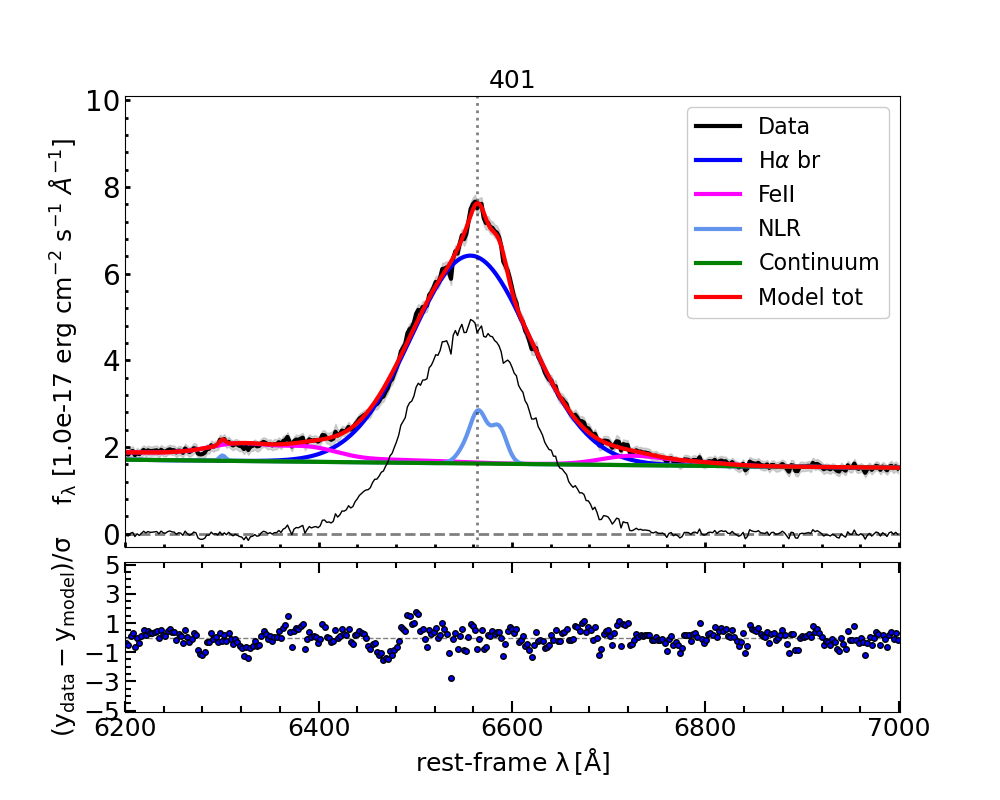}
        \includegraphics[trim = 0 0 0 0 cm, clip, width=0.38\linewidth]{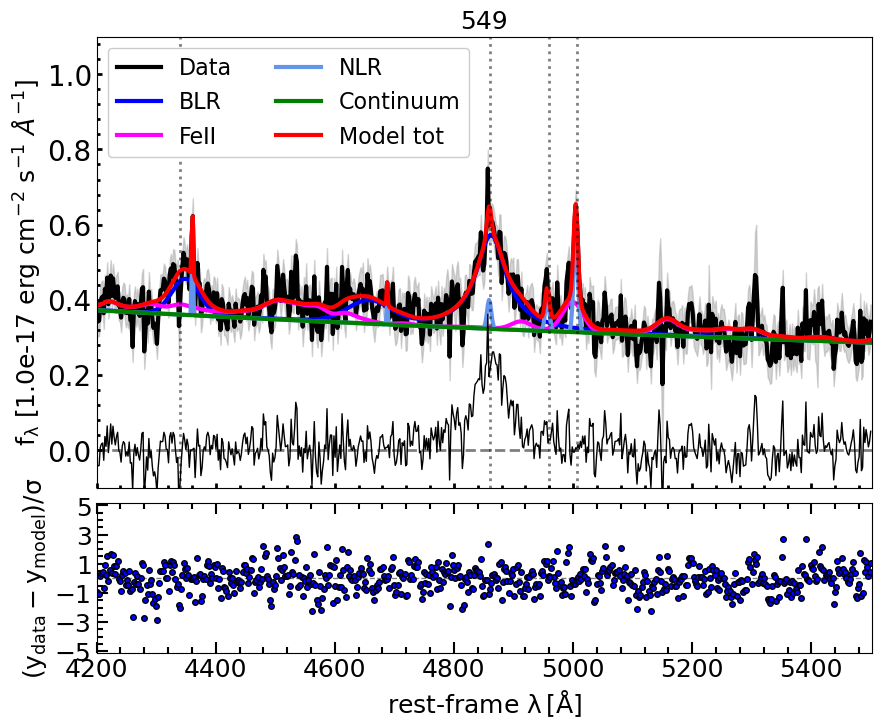}
    \includegraphics[trim = 0 0 0 1.4cm, clip, width=0.43\linewidth]{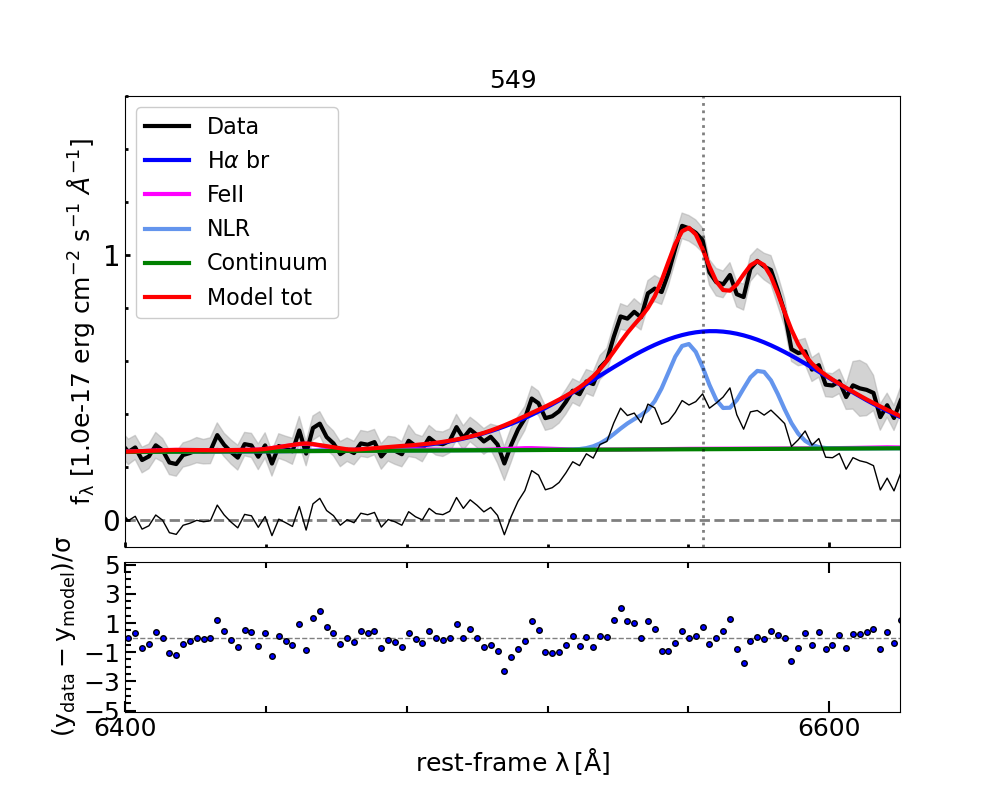}
    \caption{Same as Fig. \ref{fig:spectral_fitting} but for the targets RM032, RM312, RM401, and RM539.}
    \label{fig:spectra_1}
\end{figure*}

In this section, we show the spectrum and the best-fit modeling of the other four targeted in this work. 
Figure \ref{fig:spectra_1} shows the \hb, \oiii\ complex (left) and \ha-\nii\ complex on the right panels for each target.
The best-fit values of luminosities and FWHM for each target are reported in Table \ref{tab:best_fit_results}.

\begin{table*}[htb!]
\caption{Properties of the target and best fit results from the integrated spectra.}
\centering
\label{tab:best_fit_results}
\begin{tabular}{|c|l|l|l|c|c|c|c|c|}
\hline \hline
Target  &RA (deg)&Dec (deg) &Obs. date& $\log\left(\frac{\lambda L(5100\AA)}{\rm erg ~s^{-1}}\right)$ & $\log\left(\frac{L(H\beta}{\rm erg~s^{-1}}\right)$ & $\log\left(\frac{L(H\alpha)}{\rm erg~s^{-1}}\right)$ & \makecell{ FWHM H$\alpha$ \\ {[}km s$^{-1}${]} } & \makecell{ FWHM H$\beta$ \\ {[}km s$^{-1}${]} }  \\\hline
RM032      & 213.3063& 52.9306&May 25, 2023 & 44.94 $\pm$ 0.02                                     & 43.25    $\pm$ 0.01                           & 43.77            $\pm$ 0.01                    & 2343 $\pm$ 4                     & 2975 $\pm$ 595                  \\ \hline
RM312     & 212.4267& 52.5880&Mar 12, 2023& 44.80 $\pm$ 0.02                                     & 42.98    $\pm$ 0.07                           & 43.46 $\pm$ 0.03                                & 6611 $\pm$ 50                    & 7452 $\pm$ 121                   \\\hline
RM332     & 212.1820& 52.8280&Mar 12, 2023 & 45.16 $\pm$ 0.02                                     & 43.64   $\pm$ 0.03                            & 44.08    $\pm$ 0.01                            & 2486 $\pm$ 4                     & 4418 $\pm$ 301                   \\\hline
RM401     & 212.4886& 53.8464&Jun 2, 2023 & 45.46 $\pm$ 0.01                                      & 43.61  $\pm$ 0.01                             & 44.22   $\pm$ 0.01                             & 6615 $\pm$ 24                    & 6686 $\pm$ 75                    \\\hline
RM549     & 215.1978& 53.8000&May 1, 2023 & 44.79  $\pm$ 0.06                                   & 42.98   $\pm$ 0.09                            & 43.16 $\pm$ 0.03                               & 3530  $\pm$ 120                  & 3769 $\pm$ 387    \\  \hline            
\end{tabular}
\tablefoot{Column 1 reports the target name. Columns 2, 3, and 4 list the target coordinates and the observation date of the NIRSpec/IFU data used in this work as part of program 2057. Columns 5 to 9 report the best-fit values of the continuum monochromatic luminosity measured at 5100\AA, the broad \hb luminosity, the broad \ha luminosity, and the FWHM of the broad \ha and \hb lines, respectively. }
\end{table*}
\newpage

\section{Single-epoch relation calibrations}
\label{appendix:SE_relations}

\begin{figure}[ht!]
    \centering
    \includegraphics[width=0.8\linewidth]{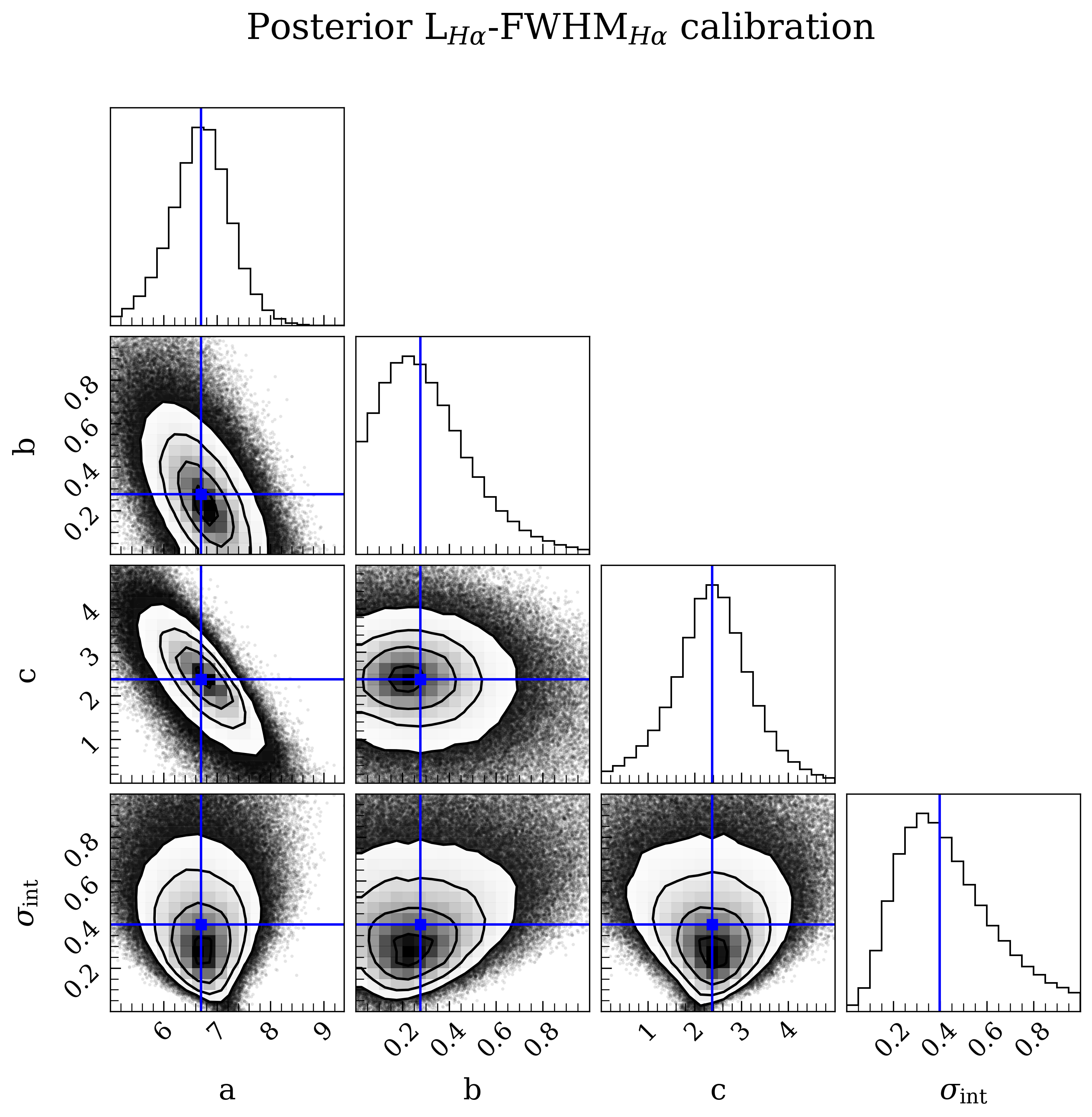}
    \caption{Posterior distributions of the parameters $a$, $b$, $c$, and $\sigma_{int}$ for the SE BH-mass estimation derived in this work by exploiting the L$_{H\alpha}$ and FWHM$_{H\alpha}$.}
    \label{fig:corner_calibrations_ha}
\end{figure}

\begin{figure}[ht!]
    \centering
    \includegraphics[width=0.8\linewidth]{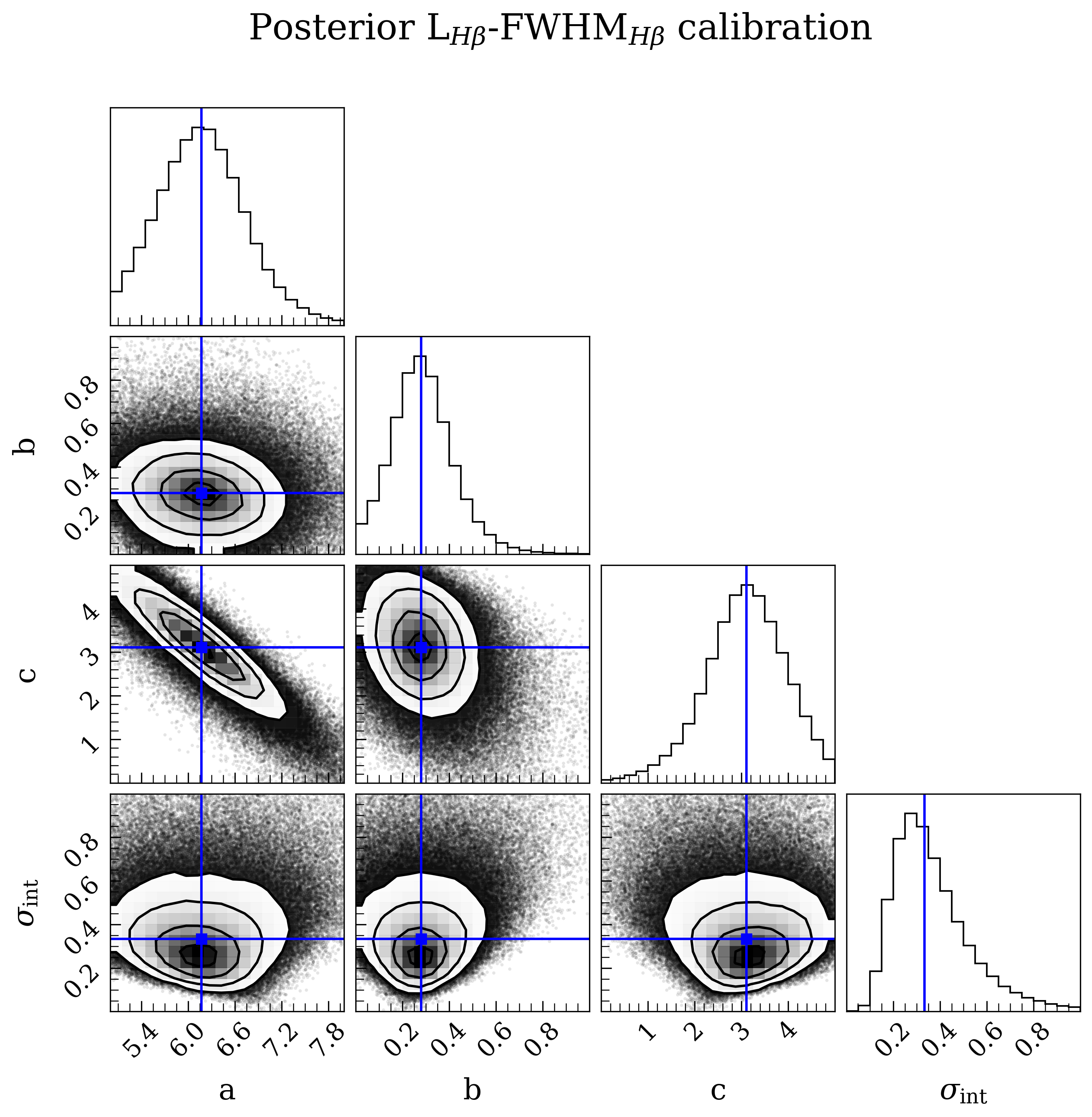}
    \caption{Same as Fig. \ref{fig:corner_calibrations_ha} but for \hb.}
    \label{fig:corner_calibrations_hb}
\end{figure}

\begin{figure}[ht!]
    \centering
    \includegraphics[width=0.8\linewidth]{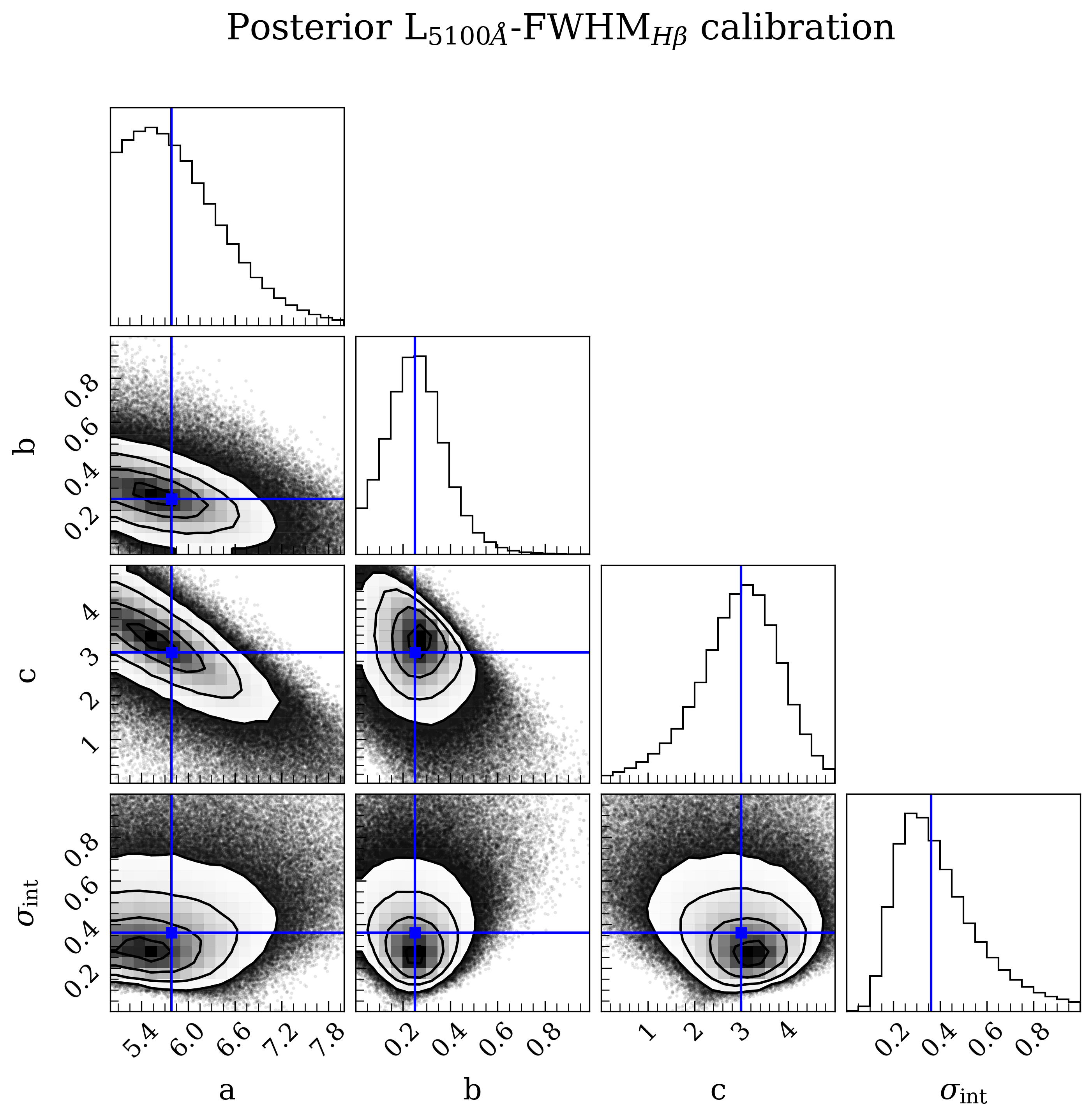}
    \caption{Same as Fig. \ref{fig:corner_calibrations_ha} but for \hb FWHM and 5100\AA luminosity.}
    \label{fig:corner_calibrations_l5100}
\end{figure}

In this section, we show the posterior distributions of the free parameters used to derive the SE scaling relation at high redshift. Figures \ref{fig:corner_calibrations_ha}, \ref{fig:corner_calibrations_hb}, and \ref{fig:corner_calibrations_l5100} display the corner plots posterior distributions for the \ha, \hb FWHM and \hb luminosity, and \hb FWHM and 5100\AA luminosity, respectively.

\newpage
\newpage
\section{Verification of our SE relations}
\label{appendix:SE_relations_test}

\begin{figure}[htb!]
    \centering
    \includegraphics[width=0.8\linewidth]{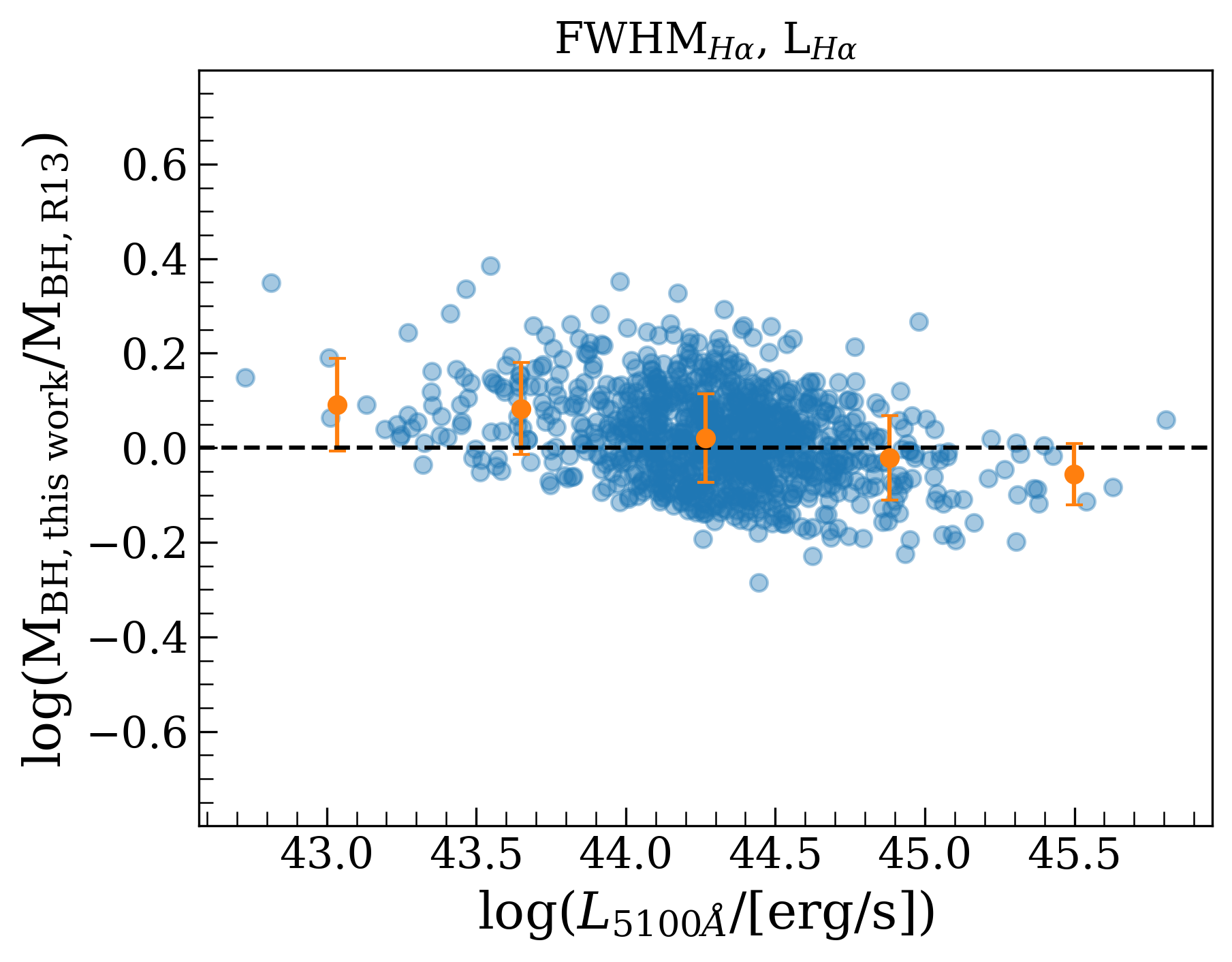}
    \caption{Difference between \Mbh inferred from our \ha-based relation and the one from \citetalias{Reines:2013} as function of the continuum luminosity.
    Blue points show the value for each QSO, while orange points show the mean and standard deviation for each luminosity bin.}
    \label{fig:delta_m_reines}
\end{figure}

\begin{figure}[htb!]
    \centering
    \includegraphics[width=0.8\linewidth]{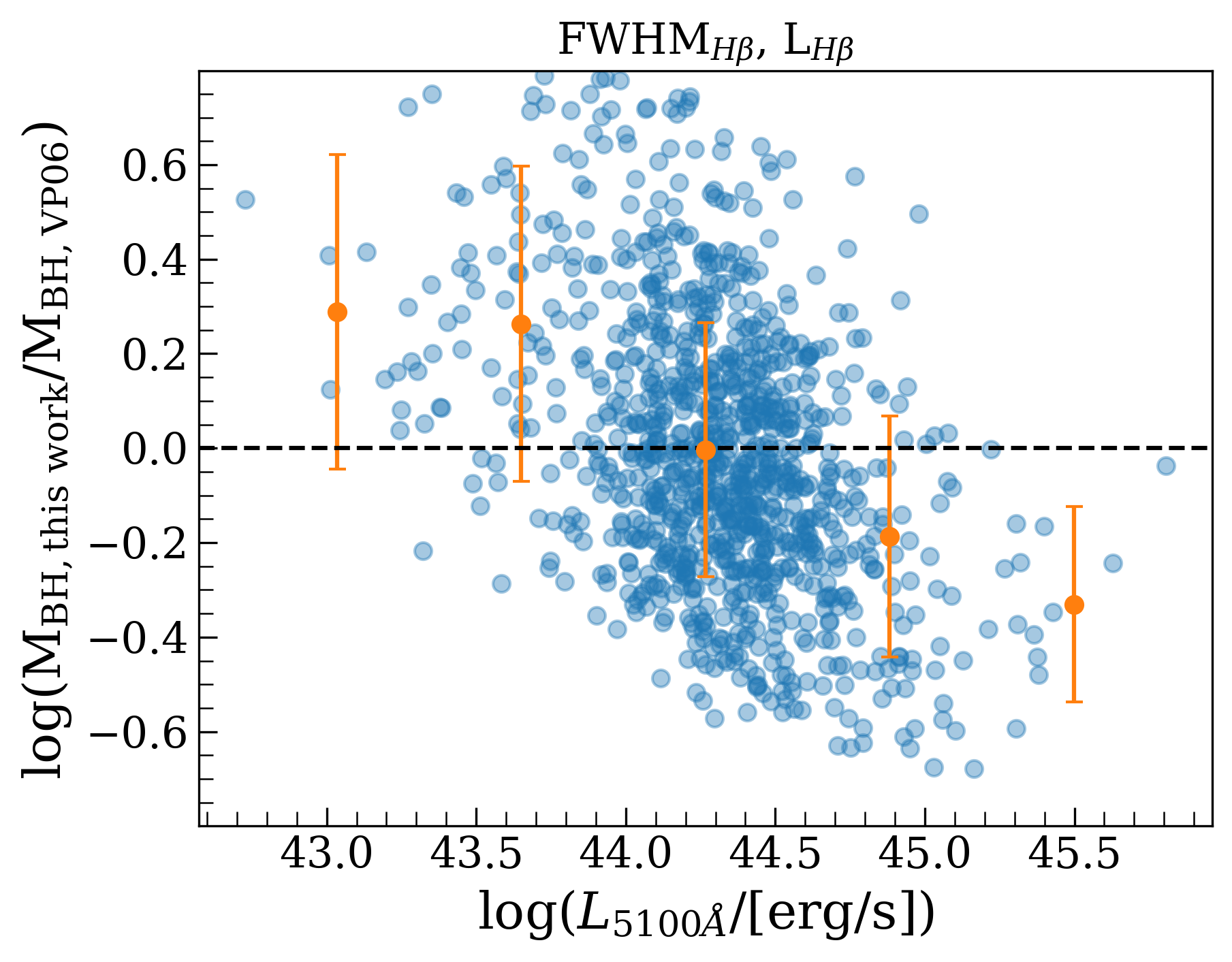}
    \includegraphics[width=0.8\linewidth]{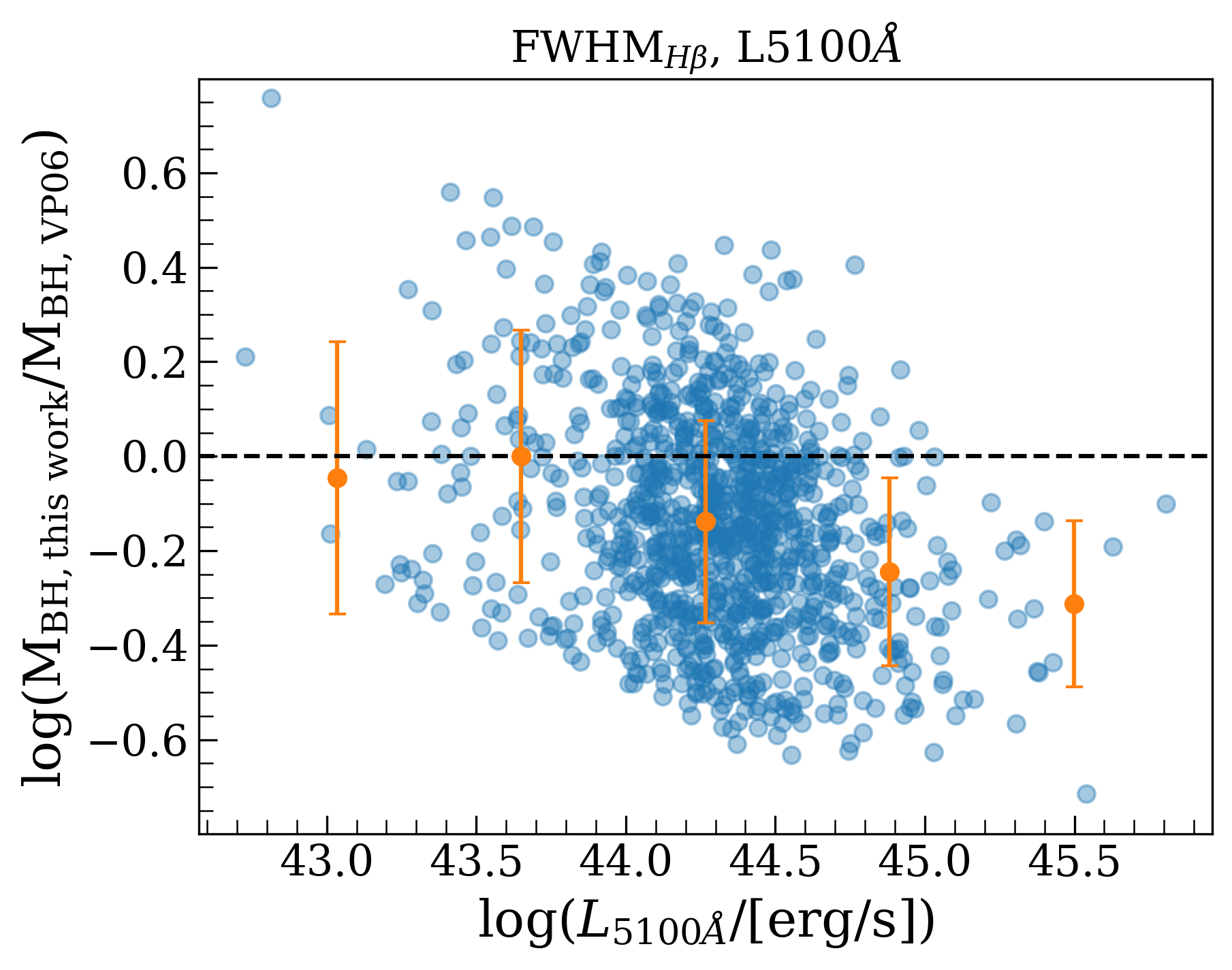}

    \caption{Same as Fig.\ref{fig:delta_m_reines} but for  \hb-based relations compared to \citetalias{Vestergaard:2006}.}
    \label{fig:delta_m_hb}
\end{figure}

In this section, we compare the \Mbh obtained with our SE relations with those from \citetalias{Vestergaard:2006} and \citetalias{Reines:2013} for a large sample of local QSOs to investigate any sign of evolution and difference.
Figures \ref{fig:delta_m_reines} and \ref{fig:delta_m_hb} show the comparison for our \ha- and \hb- based calibrators as a function of the continuum luminosity measured at 5100\AA, as a proxy for the bolometric luminosity. 
In general, we find that the \citetalias{Reines:2013} and the \citetalias{Vestergaard:2006} relations, respectively for \ha and \hb, tend to underestimate \Mbh at low luminosities ($\log(L_{5100\AA}/(\rm erg\,s^{-1}))\lesssim 45.25$) and overestimate it at high luminosities, with respect to our calibration. This effect is minor for the \ha, with the difference being on average $\lesssim$0.1 dex, whereas it is more apparent for the \hb, where the average difference reaches $\sim$0.3 dex. Obviously, a larger sample is desirable to investigate the robustness of our calibration, overcome possible biases due to the limited representativity of the QSO population, and shrink the statistical uncertainties. While on-going and future high-redshift RM campaigns will provide larger datasets to anchor SE calibrations (e.g., the ‘‘Black hole mapper'’ program; \citealt{kollmeier2017sdss}), we speculate that this behavior might be due to a physical effect, such as a flattening of the R–L relation at the high-luminosity end (e.g., \citealt{amorim2024size}).

\end{document}